\documentclass[11pt,a4paper]{article}
\usepackage{jcappub}

\usepackage{hyperref}
\usepackage{braket}
\usepackage{mathtools}
\usepackage{amsmath}
\usepackage{amsfonts}
\usepackage[dvipsnames]{xcolor}
\usepackage{graphicx}
\usepackage{caption}
\usepackage{float,subcaption}
\usepackage{amssymb}
\usepackage{tikz}
\usepackage{stackengine}
\usepackage{array}
\usepackage{booktabs,multirow}

\begin{document}

\title{Non-gravitational signals of dark energy under a gauge symmetry}
\author[a]{Kunio Kaneta,}
\emailAdd{kaneta@het.phys.sci.osaka-u.ac.jp}
\author[b]{Hye-Sung Lee,}
\emailAdd{hyesung.lee@kaist.ac.kr}
\author[b]{Jiheon Lee,}
\emailAdd{anffl0101@kaist.ac.kr}
\author[b]{Jaeok Yi}
\emailAdd{wodhr1541@kaist.ac.kr}
\affiliation[a]{Department of Physics, Osaka University, Toyonaka, Osaka 560-0043, Japan}
\affiliation[b]{Department of Physics, Korea Advanced Institute of Science and Technology, Daejeon 34141, Korea}

\abstract{
We investigate non-gravitational signals of dark energy within the framework of gauge symmetry in the dark energy sector.
Traditionally, dark energy has been primarily studied through gravitational effects within general relativity or its extensions.
On the other hand, the gauge principles have played a central role in the standard model sector and dark matter sector.
If the dark energy field operates under a gauge symmetry, it introduces the possibility of studying all major components of the present universe under the same gauge principle.
This approach marks a significant shift from conventional methodologies, offering a new avenue to explore dark energy.
}
\begin{flushright}
    OU--HET--1210
\end{flushright}
\maketitle
\flushbottom

\section{Introduction} 
\label{Sec:intro}

The Nobel prize-won discovery that the universe is in an accelerated expansion \cite{SupernovaSearchTeam:1998fmf,SupernovaCosmologyProject:1998vns} indicates the existence of the illusive component with negative pressure called dark energy. Although dark energy constitutes the majority of the energy density of the present universe, the confirmed influence of dark energy is limited to the narrow era of the recent universe through the Hubble expansion. Therefore, it remains one of the vaguest parts of cosmology, where further studies are warranted.

One of the well-known dark energy models is the quintessence field \cite{Wetterich:1987fm,Ratra:1987rm,Ferreira:1997au,Caldwell:1997ii}. The quintessence field model describes dark energy as a homogeneous scalar field rolling down the potential. It can generate an accelerated expansion in a similar way that the inflaton field generates the primordial inflation of the universe, one of the most promising scenarios for the early universe.\footnote{However, see, for instance, Ref.~\cite{Brandenberger:2009jq} for alternative scenarios. Extensive studies on the inflationary models have figured out that the inflaton can also provide the seeds for the structure formation \cite{Mukhanov:1981xt,Kodama:1984ziu} or generate the heat bath needed for the big bang nucleosynthesis (BBN) \cite{Alpher:1948ve,Hayashi:1950lqo,Walker:1991ap}.}
Reflecting the fertile physics of inflaton, we are naturally led to consider that a scalar field may exist behind the present-day expansion of the universe.

There are indeed many ultraviolet theories that imply the existence of light scalars, such as dilatons \cite{Damour:1994ya,Damour:1994zq,Gasperini:2001pc}, axions \cite{Choi:1999xn,Kim:2002tq}, and other scalar degrees of freedom in extended theories \cite{Jordan:1959eg,Brans:1961sx,Bergmann:1968ve,Wagoner:1970vr,Copeland:1997et,Bartolo:1999sq,Comelli:2003cv,Khoury:2003aq,Khoury:2003rn,Brax:2004qh,Mota:2006fz,Faulkner:2006ub,Hinterbichler:2011ca,Wang:2012kj,Brax:2012gr,Kaneta:2023rby}.
One such particle may be identified as the quintessence field.
The observation of cosmic microwave background (CMB) \cite{Planck:2018vyg} gives the present energy density of the quintessence field as tiny as $10^{-47}$ GeV$^4$, whereas the value could have been different in the early universe. The energy density of the quintessence field is determined by the field evolution throughout the thermal history of the universe.
Therefore, it could be the case that cosmological events can be affected by the quintessence field dynamics in a different way than merely introducing the cosmological constant in theories. For instance, the quintessence field may address the cosmological coincidence problem by its nontrivial dynamics \cite{Zlatev:1998tr}. 

Since the standard model (SM) is built upon gauge symmetries, it is natural to suppose that dark energy may also have a gauge symmetry. In our previous works \cite{Kaneta:2022kjj,Kaneta:2023lki}, we suggested and investigated the ``gauged quintessence’’ model where the quintessence field is charged under a new $U(1)$ gauge symmetry. Due to the interaction between the gauge boson of the new gauge symmetry and the quintessence field, this model allows rich phenomenology. We showed that the gauged quintessence model may alleviate the Hubble tension issue suffered greatly by the original quintessence field model \cite{Banerjee:2020xcn,Lee:2022cyh}, and could produce sizable relic vector density though the misalignment mechanism which is strongly suppressed in the minimal vector misalignment \cite{Nelson:2011sf}.

In this paper, we study non-gravitational signals in the gauged quintessence model. 
The dark energy sector may have a connection to the SM sector via the portals, the idea widely used in dark matter physics. Since the model has two components, that is, a vector boson and a scalar, one may use a vector portal (kinetic mixing between the hypercharge vector boson and the dark gauge boson) and/or a Higgs portal (mixing between the SM Higgs doublet and the quintessence field scalar).
We will limit ourselves to only the vector portal in this paper.
There has been discussion of non-gravitational signals of dark energy in various contexts. For instance, variation of the fundamental constants \cite{PhysRevLett.81.3067,Shaw:2005ip}, interaction to the matters \cite{Ferlito:2022mok,Vagnozzi:2021quy}, interaction to the light relic \cite{Berghaus:2020ekh,Berghaus:2023ypi},
laboratory experiments \cite{Brax:2007ak,Brax:2007hi,Homma:2019rqb,Vagnozzi:2019kvw}, and
collider \cite{Brax:2009aw,Brax:2009ey,Brax:2015hma,Brax:2016did,ATLAS:2019wdu}.
Our study differs in the sense that we use the gauge principle in studying dark energy \`a la many dark matter studies using the same principle.
We will demonstrate how dark energy can influence the production of the dark gauge boson, and induce various observational astrophysical signals.
 
This paper is organized as follows. In Sec.~\ref{Sec:gq}, we introduce the basic formalism of the gauged quintessence model with the kinetic mixing. In Sec.~\ref{sec:pddgb}, we discuss the production and decay of the dark gauge boson through kinetic mixing, and how they differ from the conventional dark photon model. In the subsequent sections, we discuss the constraints on our scenario from overproduction of the dark gauge boson, CMB distortion, diffuse X-ray/gamma-ray background, and distinguishable non-gravitational signals. We summarize and conclude in Sec.~\ref{Sec:summary}.

\section{Gauged quintessence model with kinetic mixing}
\label{Sec:gq}

The gauged quintessence model \cite{Kaneta:2022kjj} includes a complex scalar $\Phi$ and a $U(1)_\text{dark}$ gauge boson $\hat{X}_\mu$. The complex scalar is charged under $U(1)_\text{dark}$ gauge symmetry, the radial part of $\Phi$ is taken as the quintessence field ($\phi$), and the dark gauge boson gets the mass proportional to $\phi$. As the quintessence field $\phi$ value evolves, $\hat X$ has a mass-varying characteristic. (For some other examples of the mass-varying particles, see Refs.~\cite{Casas:1991ky,Garcia-Bellido:1992xlz,Anderson:1997un,Fardon:2003eh,Berlin:2016bdv,Krnjaic:2017zlz,Davoudiasl:2019xeb,Boubekeur:2023fqo,ChoeJo:2023ffp,ChoeJo:2023cnx}.)

The dark gauge boson and SM hypercharge boson ($\hat{B}_{\mu}$) can be mixed via kinetic mixing \cite{Holdom:1985ag}. Hence, the action for the gauge field is written as\footnote{The FLRW metric with $g_{\mu\nu} = (-1,a^2,a^2,a^2)$ is used.}
\begin{equation}
    \mathcal{L}_{\text{gauge}} \supset -\frac{1}{4}\hat{B}_{\mu\nu}\hat{B}^{\mu\nu}+\frac{\varepsilon}{2 \cos\theta_W}\hat{B}_{\mu\nu}\hat{X}^{\mu\nu}-\frac{1}{4}\hat{X}_{\mu\nu}\hat{X}^{\mu\nu} -\frac{1}{2}(g_X^{}\phi)^2 \hat{X}_{\mu}\hat{X}^{\mu},
\end{equation}
where $\varepsilon$ is a kinetic mixing, $g_X$ is $U(1)_{\text{dark}}$ gauge coupling, and $\theta_W$ is the weak mixing angle. Due to the kinetic mixing, the kinetic terms of the gauge fields are not diagonal. They are diagonalized by some rotations among the gauge boson fields:
\begin{equation}
    \mathcal{L}_{\text{gauge}} \supset -\frac{1}{4}F_{\mu\nu}F^{\mu\nu}-\frac{1}{4}X_{\mu\nu}X^{\mu\nu} -\frac{1}{2}(\eta g_X^{}\phi)^2 X_{\mu}X^{\mu},
    \label{eq:diagLagrangian}
\end{equation}
where $F_{\mu\nu}$ is electromagnetic field strength tensor, $X_\mu$ and $X_{\mu\nu}$ are diagonalized dark gauge boson field and its strength tensor and $\eta \equiv 1/\sqrt{1-\varepsilon^2/\cos^2\theta_W}$. (See App.~\ref{sec:Diagonalization of kinetic and mass terms} for the detail.) Since $\varepsilon$ is typically constrained by various experiments and observations, we always take it as a small parameter, and approximate $m_X  \approx g_X\phi$. On this basis, $X$ couples to the SM fermions, and such couplings are proportional to $\varepsilon$.

Although the photon is massless, it can obtain a non-zero effective mass through the interaction with particles in the thermal bath. For instance, Compton scattering with the electrons gives the real part of effective photon mass as \cite{Altherr:1992mf,Braaten:1993jw}
 \begin{equation}
     m_{\gamma}^2= \begin{cases}
         4\pi \alpha_\mathrm{em} (n_e/m_e) \quad &\text{for} \quad T \ll m_e, \\
         (2/3)\pi\alpha_\mathrm{em} T^2 \quad &\text{for} \quad T \gg m_e,
     \end{cases}
 \end{equation}
where $\alpha_\mathrm{em}\approx1/137$ is the fine structure constant and $T$ is the temperature of the thermal bath. In the presence of the kinetic mixing, thermal effects also generate a non-diagonal mass term between the photon and dark gauge boson. In a basis where the photon and dark gauge boson mass terms are diagonal,\footnote{Due to the additional diagonalization, the mass of dark gauge boson in the medium and vacuum has a tiny difference proportional to $\varepsilon$, which we can neglect.} the effective kinetic mixing is given as \cite{Redondo:2008aa}
\begin{equation}
     |\overline{\varepsilon}(\omega,T)|^2 = \varepsilon^2 \frac{m_X^4}{(m_X^2-m_{\gamma}^2)^2+(\omega D)^2},
\end{equation}
where $\omega D$ is the imaginary part of the effective photon mass \cite{Redondo:2008ec}.\footnote{$D \sim 8\pi\alpha_{\mathrm{em}}^2/(3m_e^2)n_e$ when $T\ll m_e$, and $\omega \sim T$.} This results in the kinetic mixing being suppressed in $m_X \ll m_{\gamma}$ limit, while there is a resonance at $m_{X} = m_{\gamma}$. In the resonance regime, the kinetic mixing is amplified as $|\overline{\varepsilon}|=\varepsilon m_X^2/(\omega D)$.

Let us now discuss the dynamics of the model. As shown in Eq.~\eqref{eq:diagLagrangian}, the mass of the dark gauge boson is proportional to the quintessence field value, which can be determined from its equation of motion \cite{Kaneta:2022kjj},
\begin{equation}
\begin{split}
    &\ddot{\phi} +3H\dot{\phi} +\frac{\partial V_{\mathrm{eff}}(\phi)}{\partial\phi}  =0,\\
    &V_{\text{eff}}(\phi)= V_{0}(\phi)+ \frac{1}{2}g_X^2X_{\mu}X^{\mu}\phi^2,
\end{split}
\label{eq:Qeom}
\end{equation}
where $V_0(\phi)$ is the potential of $\phi$, and $g_X^2X_{\mu}X^{\mu}\phi^2/2$ is the gauge potential \cite{Kaneta:2022kjj} which gives an additional contribution to the quintessence potential. While there is a freedom to choose $V_{0}(\phi)$ as long as it gives the required dark energy phenomenology,\footnote{For the uncoupled quintessence field model, the dark energy phenomenology is obtained when $ \sqrt{\partial^2V_0/\partial\phi^2} \lesssim H$. } we use the inverse-power potential suggested by Ratra and Peebles \cite{Ratra:1987rm} throughout this paper:
\begin{equation}
V_0 (\phi) = \frac{M^{\alpha+4}}{\phi^{\alpha}} \, ,
\label{eq:RPpotential}
\end{equation}
where $\alpha>0$, and $M$ is chosen to fit the present-time dark energy density. In this work, we take $V_0(\phi)$ as a quantum effective potential, which includes all the quantum corrections.\footnote{Such corrections include vacuum energy of all field configurations and $\phi$ dependent correction from the dark gauge boson. We presume that there exists a UV theory that passes down the Ratra-Peebles potential as an IR effective potential regardless of the complication of the UV tree level potential.} This allows us to consider relatively larger $g_X$ and $m_X$ compared to our previous works \cite{Kaneta:2022kjj,Kaneta:2023lki}.\footnote{Possible constraints on $g_X$ and $m_X$ when taking the Ratra-Peebles potential as a tree-level potential are given in Ref.~\cite{Kaneta:2022kjj}.} 

The Ratra-Peebles potential has a tracking behavior, so the wide range of initial conditions eventually converge to one common tracking solution \cite{Steinhardt:1999nw} given by \cite{Zlatev:1998tr}
\begin{equation}
\begin{split}
   & \phi \propto a^{3(1+w_b)/(2+\alpha)},\\ &\sqrt{\frac{\partial^2V_0(\phi)}{\partial\phi^2}}\sim 3 H,
    \label{eq:tracksol}
\end{split}
\end{equation}
where $w_b$ is the equation-of-state parameter of the background component of the universe,\footnote{$w_b=1/3$ for the radiation-dominated era, and $w_b=0$ for the matter-dominated era.} and the second line (which would be a quintessence field mass without the gauge potential) can be derived from Eq.~\eqref{eq:Qeom} with $X^{\mu}=0$. However, the tracking solution of the $\phi$ field can be interrupted by the backreaction of the background dark gauge boson. If $V_{\text{eff}}(\phi)$ is steep around the potential minimum such that $m_\phi = \sqrt{\partial^2V_{\text{eff}}/\partial\phi^2} \gg H$, $\phi$ field is trapped at the minimum of $V_{\text{eff}}$ \cite{Kaneta:2023lki}. In this case, we approximate $\phi \sim \phi_{\mathrm{min}}$, where $\phi_{\mathrm{min}}$ is the value of $\phi$ at the potential minimum. 

 An explicit form of the backreaction can be computed from the thermal average of $ \langle X_{\mu}X^{\mu}\rangle $ as \cite{Linde:1978px}
\begin{equation}
    \langle X_{\mu}X^{\mu}\rangle = 3 \int \frac{d^3\vec{p}}{(2\pi)^3} \frac{f(\vec{p})}{\sqrt{m_X^2+|\vec{p}|^2}},
    \label{eq:XX}
\end{equation}
where $f(\vec{p})$ is the phase space distribution of the dark gauge boson. This expression can be simplified if one assumes that the phase space distribution is isotropic and peaked at a certain momentum scale $|\vec{p}|\sim T $. If the dark gauge bosons are relativistic, $T\gg m_X$,
\begin{equation}
    \langle X_{\mu}X^{\mu} \rangle \approx C \frac{n^{}_X}{T},
    \label{eq:XXrel}
\end{equation}
where $n_X$ is the number density of the dark gauge boson, and $C$ is the correction factor, which depends on the actual distribution. If the dark gauge boson follows the Bose-Einstein distribution, we found that $C= 0.68$. On the other hand, if the dark gauge bosons are non-relativistic, $T \ll m_X$, then 
\begin{equation}
    \langle X_{\mu}X^{\mu}\rangle \approx \frac{n^{}_X}{m_X^{}}.
    \label{eq:XXnonrel}
\end{equation}
Then, the condition for the trap of $\phi$ is written as
\begin{equation}
    n_X^{} \gg \begin{dcases}
          \frac{H^2 T}{(\alpha+2)C g_X^2}\quad &\mathrm{for}\; T\gg m_X,
         \\\frac{H^2 m_X}{(\alpha+2) g_X^2} \quad &\mathrm{for}\; T\ll m_X.
    \end{dcases}
    \label{eq:trapCondition}
\end{equation}
$\phi$ is easier to be trapped for a larger $g_X$. Also, this expression shows that
even if $\phi$ is once trapped, it could be released to the tracking solution as $n_X$ decays away or $m_X$ increases.

One can also calculate $\phi_{\textrm{min}}$ using Eq.~\eqref{eq:XXrel} or \eqref{eq:XXnonrel}, then $m_X$ is given as follows: 
\begin{equation}
    m_X^{} \sim \begin{dcases}
         &\left(\frac{ \alpha g_X^{\alpha} M^{4+\alpha}}{ C \dfrac{n^{}_X}{T}}\right)^{\frac{1}{\alpha+2}} \quad \mathrm{for}\; T\gg m_X,
         \\&\left(\frac{\alpha g_X^{\alpha} M^{4+\alpha}}{n_X^{}}\right)^{\frac{1}{\alpha+1}} \quad \mathrm{for}\; T\ll m_X.
         \end{dcases}
          \quad (\mathrm{trapped})
         \label{eq:mXtrap}
\end{equation}
When the condition of Eq.~\eqref{eq:trapCondition} is satisfied, we use Eq.~\eqref{eq:mXtrap} to find $m_X$. In other case, we assume that $\phi$ follows the tracking solution, then $m_X^{}$ is given from the second line of Eq.~\eqref{eq:tracksol} as follows:
\begin{equation}
    m_X \sim g_X^{} \left( \frac{\sqrt{\alpha(\alpha+1)M^{4+\alpha}}}{3 H}\right)^{\frac{2}{\alpha+2}}. \quad (\mathrm{tracking})
    \label{eq:mXtrack}
\end{equation}

We presented a general formalism applicable to any values of $\alpha > 0$, and the following discussions are largely valid regardless of any $\alpha > 0$. In our quantitative analysis, including all figures, we will set $\alpha = 1$, for definiteness.

\section{Production and decay of the dark gauge boson}\label{sec:pddgb}

The dark gauge boson can be produced from the SM thermal bath via kinetic mixing.\footnote{For an alternative production using the dark axion portal, see Refs.~\cite{Kaneta:2016wvf,Kaneta:2017wfh}.} Since the effective kinetic mixing in the thermal bath largely depends on the mass of the dark gauge boson, it is crucial to incorporate a mass-varying effect in the production process. Also, the background dark gauge boson density can affect the evolution of the dark gauge boson mass, and we will demonstrate the self-consistent calculation of dark gauge boson density.

We investigate the production of a light dark gauge boson ($m_X^{} \ll m_e$), which becomes heavier in the late universe due to the mass-varying nature.\footnote{If the dark gauge boson is heavier than $2m_e$, the dominant production channel is the pair coalescence ($e^+e^- \rightarrow X$) \cite{Redondo:2008ec}. However, there is an upper limit to the production from this channel since sufficient background dark gauge boson would suppress $m_X$ to be smaller than $2m_e$. (One can get the upper limit from Eq.~\eqref{eq:mXtrap} by substituting $2m_e$ for $m_X$.) We checked that the contribution from the coalescence is negligibly small in the parameter space we consider in this paper.} For the light dark gauge boson, Compton-like process ($\gamma e^{-} \rightarrow X e^{-}$) is the dominant production channel \cite{Redondo:2008ec}. Assuming that backreaction from $X$ to SM thermal bath is negligible, or produced $n_X$ is small compared to the SM thermal bath entropy density $s$, the Boltzmann equation is given as \cite{EscuderoAbenza:2020cmq}
\begin{equation}
    \frac{dn_X^{}}{dt}+3Hn^{}_X= \frac{1}{2\pi^4} \int ^{\infty}_{S_0} dS\; p_{12}^2 \sqrt{S} \sigma_{\text{Comp}}(S) \left( T e^{\mu/T} K_1\left(\frac{\sqrt{S}}{T}\right) \right),
    \label{eq:boltz}
\end{equation}
where $S_0=(m_e+m_{\gamma})^2$, $p_{12}=\sqrt{S-(m_e+m_{\gamma})^2}\sqrt{S-(m_e-m_{\gamma})^2}/(2\sqrt{S})$, $\mu$ is the chemical potential of the electron, and $T$ is the temperature of SM thermal bath. 
The effect of $\mu$ becomes significant below $e^+e^-$ annihilation around $T\sim \mathcal{O}(10)$ keV \cite{Thomas:2019ran}.
Also, $\sigma_{\text{Comp}}$ is the Compton scattering cross-section given as \cite{Redondo:2008ec}
\begin{multline}
    \sigma_\text{Comp} (S) = \frac{2\pi\alpha_{\mathrm{em}}^2 \overline\varepsilon^2}{(S-m_e^2)^3} \bigg(\frac{\beta}{2S}\big(S^3+15S^2 m_e^2 - S m_e^4 +m_e^6 + m_X^2 (7S^2 + 2S m_e^2-m_e^4)\big)\\+2 (S^2 - 6S m_e^2 - 2 m_e^4 - 2m_X^2(S-m_e^2 - m_X^2))\log\left[\frac{S(1+\beta)+m_e^2 - \mu^2}{2m_e\sqrt S}\right] \bigg),
\end{multline}
where $S^2\beta^2 = (S-(m_e+m_X)^2)(S-(m_e-m_X)^2)\approx (S-m_e^2)^2$.

As discussed in Ref.~\cite{Redondo:2008ec}, the dark gauge boson is resonantly produced when $m_\gamma \approx m_X$. Since $m_X=g_X \phi$, the value of $\phi$ determines when the resonant production occurs. If there is no sufficient background density of $X$, then $\phi$ follows the tracking solution. We assume that the initial abundance of the dark gauge boson is negligibly small, and $\phi$ field initially follows the tracking solution, where the mass is given from Eq.~\eqref{eq:mXtrack}. If the sufficient dark gauge boson is produced before the resonance, the $\phi$ field follows the trapped solution, and its mass is given from Eq.~\eqref{eq:mXtrap}. For given parameters $\varepsilon$ and $g_X$, we numerically solved the Boltzmann equation, Eq. \eqref{eq:boltz}, with the tracking mass given in Eq. \eqref{eq:mXtrack} and checked whether a sufficient amount of $X$ is produced for $\phi$ to be trapped. Once the trapping occurred, we checked when it happened, and from that time, we solved Eq. \eqref{eq:boltz} with the trapped mass in Eq. \eqref{eq:mXtrap}. Based on this analysis, we can check the resonant behavior of the production.

\begin{figure}[t]
\centering
\begin{subfigure}{.48\textwidth}
\centering\includegraphics[width=\linewidth]{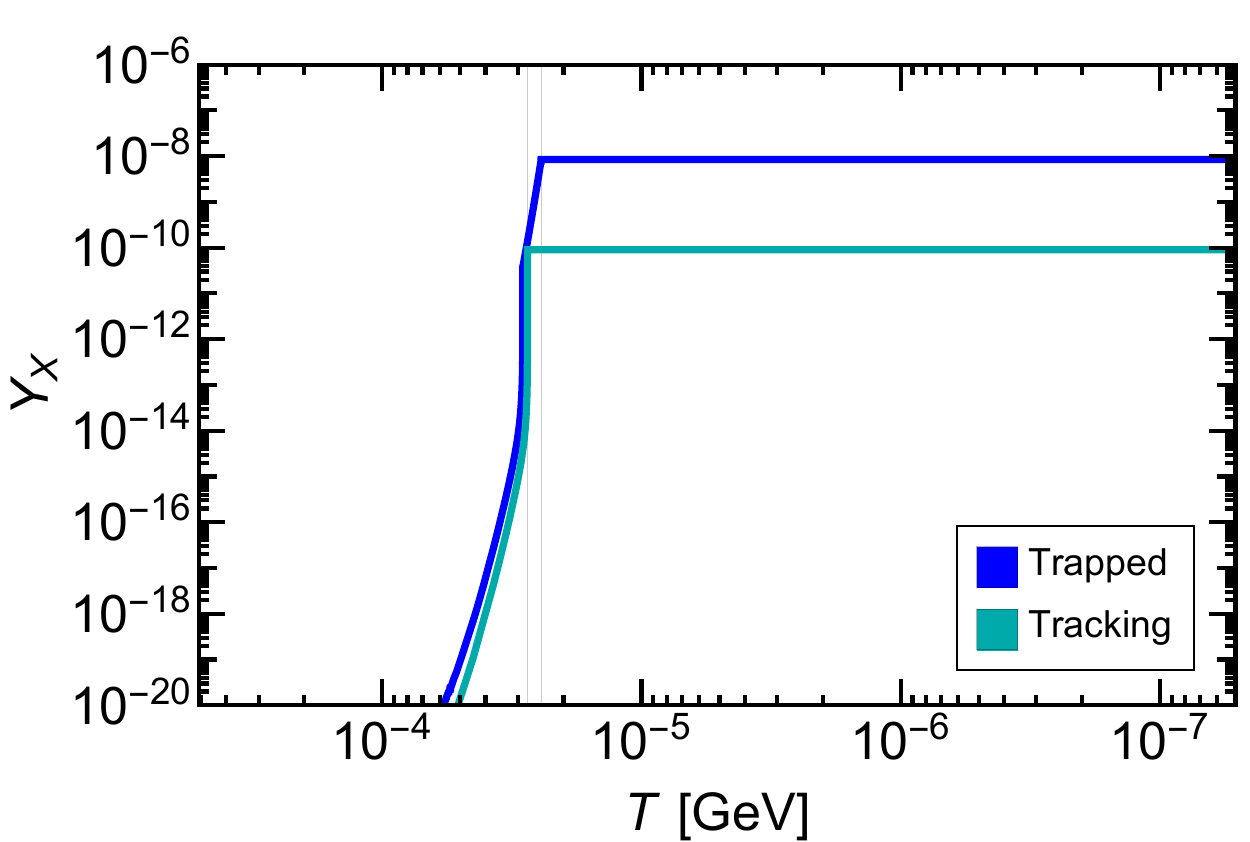}
\caption{$Y_X$}\label{fig:evden}
\end{subfigure}  \quad
\begin{subfigure}{.48\textwidth}
\centering\includegraphics[width=\linewidth]{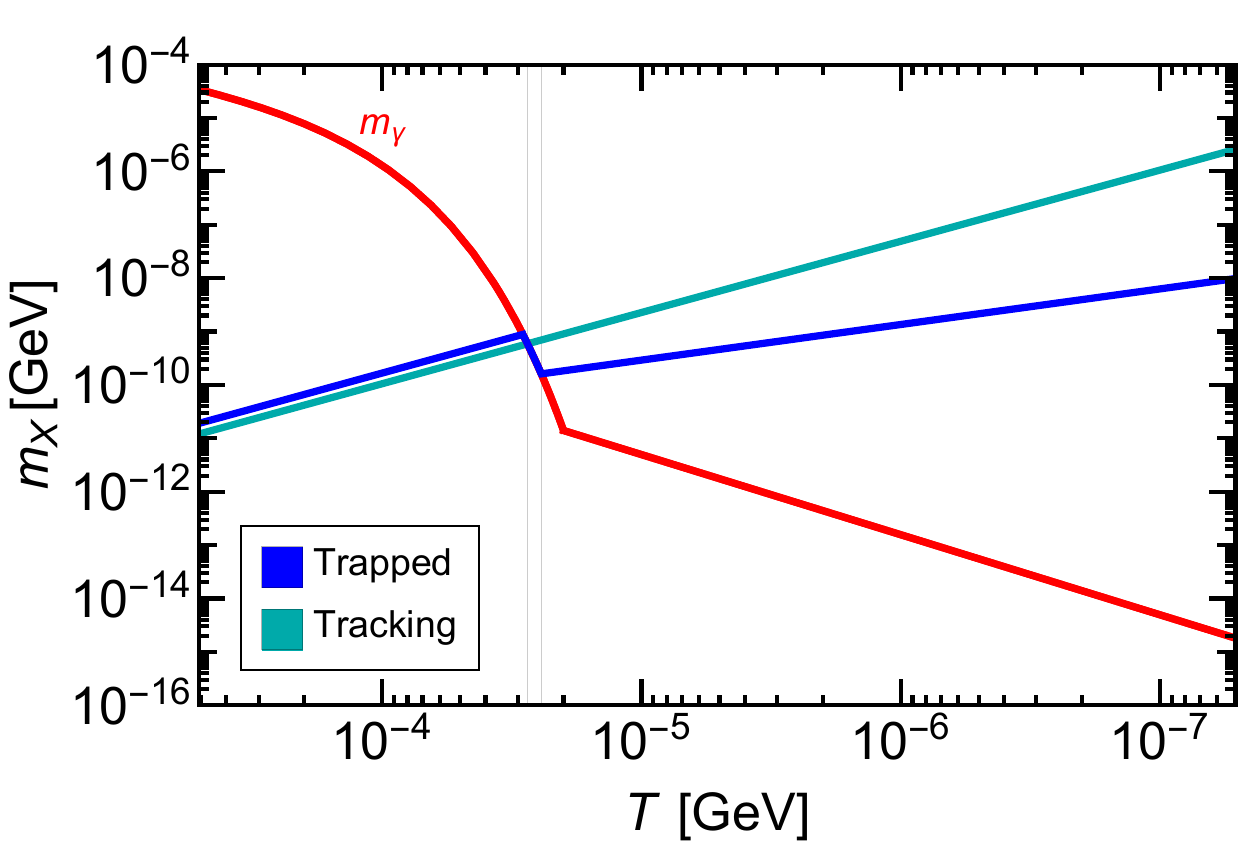}
\caption{$m_X$ and $m_\gamma$}\label{fig:evmass}
\end{subfigure}
\caption{Examples of the evolution of $Y_X\equiv n_X/s$ and $m_X$ around resonant production. The blue curve corresponds to the trapped production ($g_X = 5 \times 10^{-19}$). The cyan curve corresponds to the tracking production ($g_X = 3\times 10^{-19}$). Both scenarios share $\varepsilon=10^{-11}$ in common.  The thin vertical lines correspond to when the resonant production finishes. The red curve in (b) corresponds to the effective photon mass. The production of the trapped case is much larger compared to the tracking case because the resonance period is extended. (We set $\alpha=1$ throughout this work.)}
\label{fig:ev}
\end{figure}

The resonant behavior depends on whether $\phi$ is trapped or following the tracking solution when the resonance occurs. If $\phi$ follows the tracking solution, $m_X$ monotonically increases over time, while $m_\gamma$ decreases over time. So, there is an instantaneous moment when $m_X$ and $m_\gamma$ coincide, and the resonance occurs in an instant. In the case of the trapped $\phi$, however, the resonant production of $X$ suppresses $m_X^{}$ [Eq.~\eqref{eq:mXtrap}]. If the production is sufficiently large to keep $m_X^{}$ from surpassing $m_\gamma$, then $m_X^{}$ could follow $m_{\gamma}$ during some period. Therefore, the resonance can be sustained until the production rate becomes too weak due to the dilution of the photon and electron by the expansion of the universe. Figure~\ref{fig:ev} shows examples of the production for each case. The blue (cyan) curve corresponds to the trapped (tracking) case, respectively. In Fig.~\ref{fig:evden}, the number density of $X$ normalized by the SM thermal bath entropy density $s$ ($Y_X \equiv n_X/s$), drastically increases at the resonance, i.e. when $m_X = m_\gamma$ holds. Also, Fig.~\ref{fig:evmass} shows that $m_X$ of the trapped case (blue curve) follows $m_{\gamma}$ (red curve) during some interval. This extended resonance clearly enhances $Y_X$ by several orders of magnitude compared to the tracking case.

\begin{figure}[t]
    \begin{subfigure}{.5\textwidth}
    \centering
    \includegraphics[width=\linewidth]{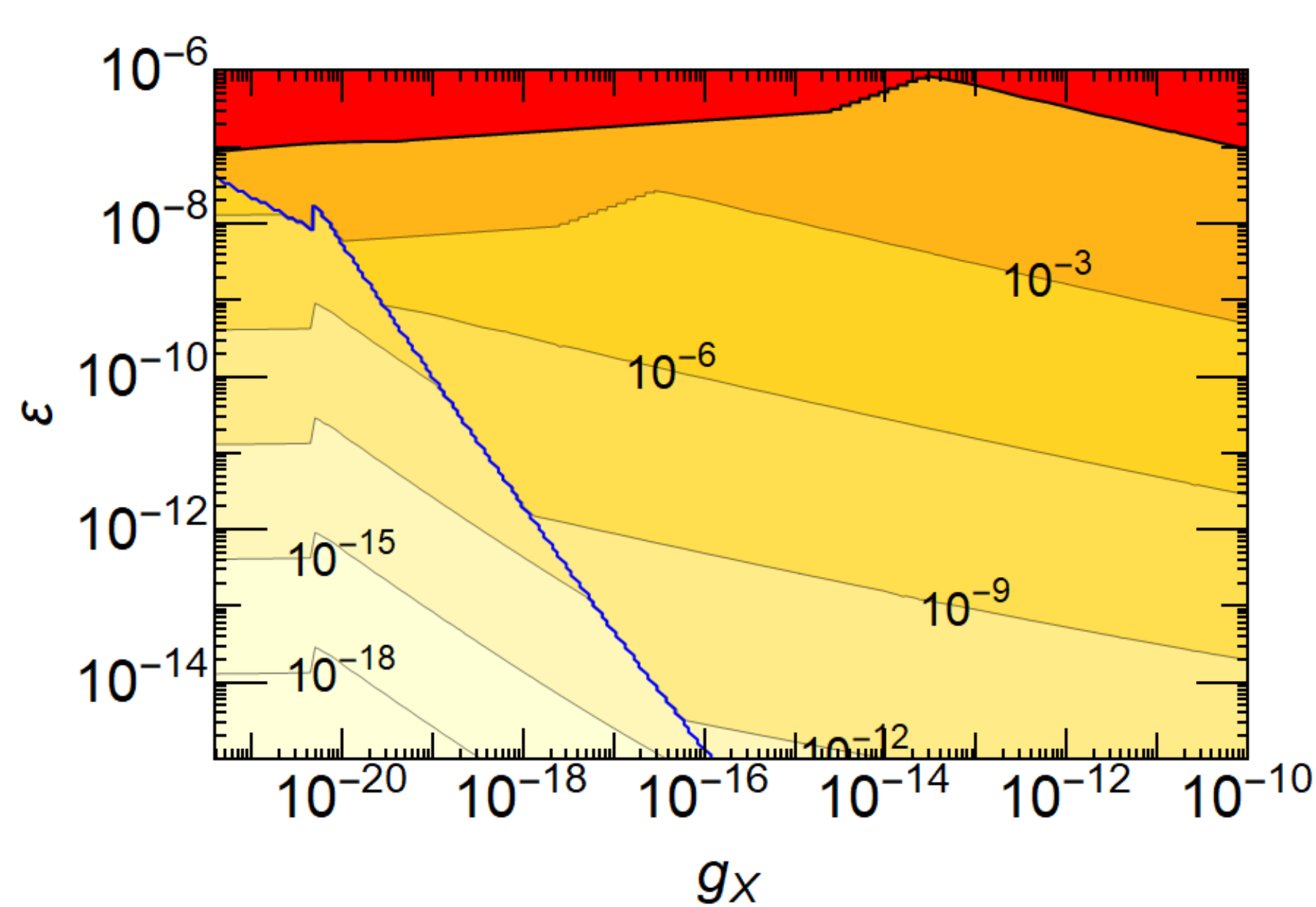}  \caption{$Y_X$}  \label{fig:nprod}\end{subfigure}\begin{subfigure}{.5\textwidth}\centering \includegraphics[width=\linewidth]{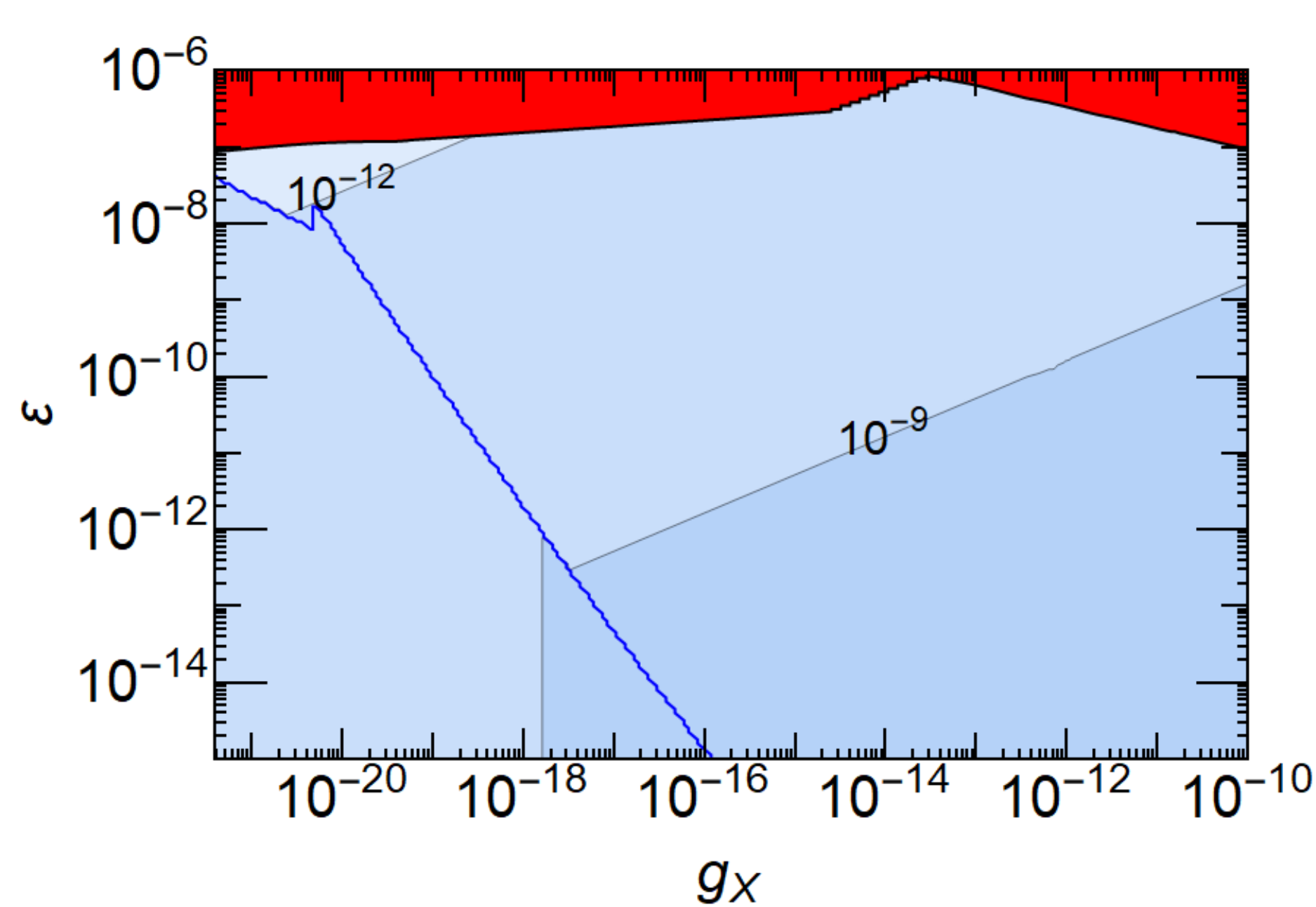}  
    \caption{$m_X$ [GeV]}  \label{fig:mprod}
    \end{subfigure}
    \caption{The dark gauge boson density normalized by the entropy density of SM thermal bath ($Y_X\equiv n_X/s$) and the mass ($m_X$) when they are resonantly produced. The red region corresponds to the parameter space that gives $Y_X\ge 1$. The blue curve is the boundary between the tracking (below) and trapped (above) cases. } 
    \label{fig:prod}
\end{figure}

The resultant $Y_X$, for each $\varepsilon$ and $g_X$, when $X$ is produced is given in Fig.~\ref{fig:nprod}. The parameter space below the blue curve corresponds to the tracking production, and the other side corresponds to the trapped production. The production of $X$ is amplified in the trapped case, as the resonance is extended. Also, Fig.~\ref{fig:mprod} shows the mass of the dark gauge boson when it is produced. Both the density and mass of the dark gauge boson change significantly across the blue curve. The contour lines below the blue curve in Fig.~\ref{fig:mprod} are vertical since the tracking solution does not depend on $Y_X$ ($n_X$) [Eq.~\eqref{eq:mXtrack}]. 

After the resonant production, there are two scenarios for the dark gauge bosons, depending on whether they (i) decay into other components, or (ii) remain as the sub-dominant mass-varying dark matter. The main decay channels of the dark gauge boson are $X\rightarrow \gamma\gamma\gamma$.\footnote{The dark gauge bosons also decay to neutrinos ($X\rightarrow \nu\bar{\nu}$), and this is the dominant decay channel if the dark gauge boson is lighter than $\mathcal{O}(10)$ $\text{keV}$ \cite{Ibe:2019gpv}. However, $X$ density is hardly affected by the neutrino decay even when the neutrino channel is the dominant one, since the decay rate into the neutrinos is extremely small. For instance, with the most promising set of parameter $\varepsilon = 10^{-7}$ and $m_X^{}=10$ keV, one obtains $\Gamma_{\nu\bar{\nu}}/H_0 \sim 10^{-10}$.}, and $X\rightarrow e^+e^-$ processes. The decay products (three photons, and $e^+ e^-$ pair) might become the main signal of the gauged quintessence model. 

\begin{figure}
    \centering
    \includegraphics[width=.8\textwidth]{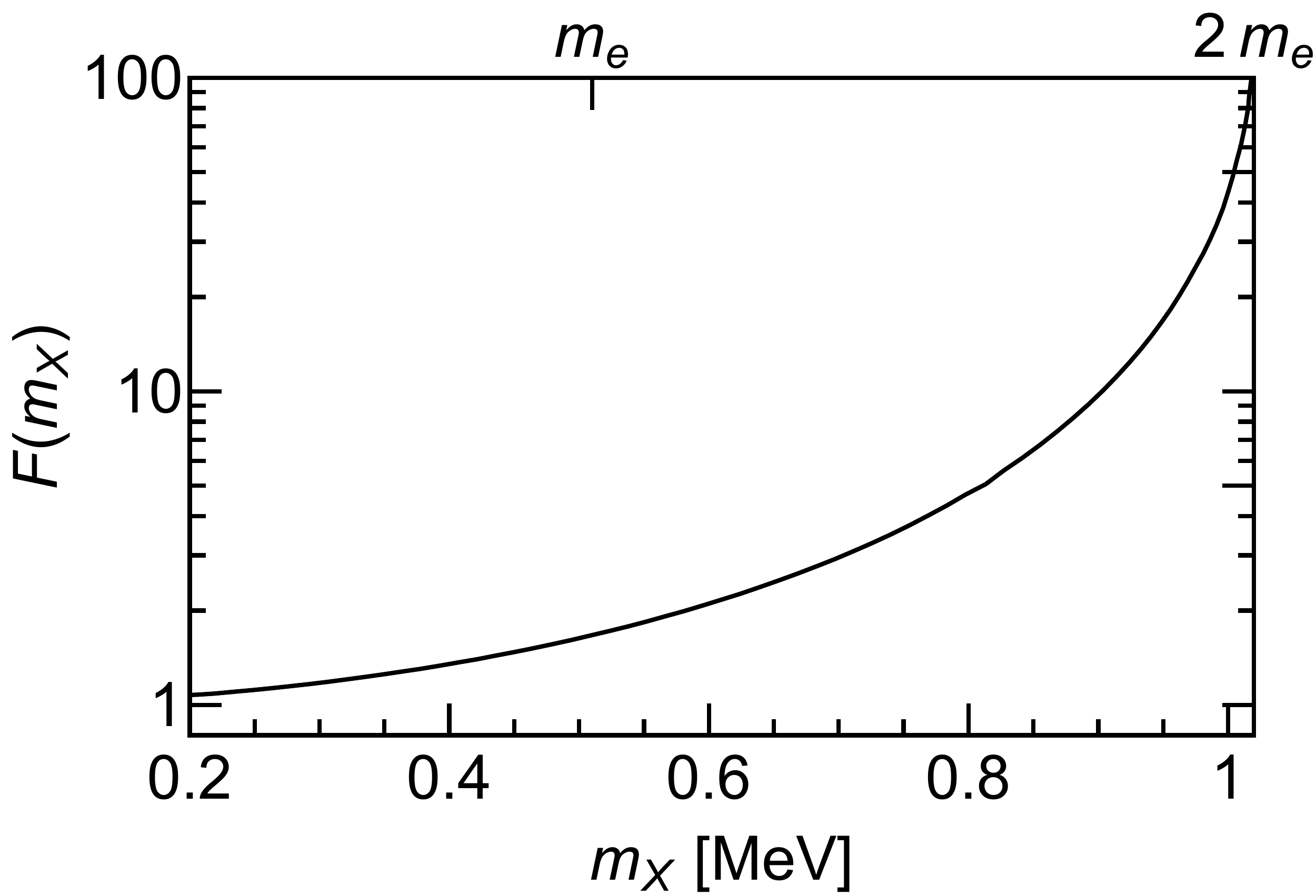}
    \caption{Enhancement factor $F(m_X)$ for three-photon decays \cite{McDermott:2017qcg}.}
    \label{fig:enhf}
\end{figure}

Figure~\ref{fig:mprod} shows that initially, the mass of the dark gauge boson was smaller than $2m_e$ in the parameter space of interest. Therefore, in the early universe, the $X\rightarrow e^+e^-$ channel is closed, but the $X\rightarrow \gamma\gamma\gamma$ process can occur. $X\rightarrow\gamma\gamma\gamma$ decay rate below $e^+e^-$ threshold ($m_X < 2m_e$) is given by \cite{Pospelov:2008jk,McDermott:2017qcg}
\begin{equation}
    \Gamma_{\gamma\gamma\gamma} =  F(m_X^{})\frac{17\alpha_{\mathrm{em}}^4\overline{\varepsilon}^2 }{11664000\pi^3}\frac{m_X^9}{m_e^8} ,
    \label{eq:3photondecayrate}
\end{equation}
where $F(m_X)$ is the enhancement factor, which leads to a substantial effect when $m_X$ is close to $2 m_e$. $F(m_X)$ is described in Fig.~\ref{fig:enhf} \cite{McDermott:2017qcg}. If $\Gamma_{\gamma\gamma\gamma}$ becomes larger than Hubble parameter $H$, most of the dark gauge boson energy density is converted to the photon energy density.

As the mass of the dark gauge boson grows according to either Eq.~\eqref{eq:mXtrap} or Eq.~\eqref{eq:mXtrack}, $X\rightarrow e^+e^-$ channel can open when the mass of the dark gauge boson crosses $e^+e^-$ threshold. $e^+e^-$ decay rate is given by
\begin{equation}
    \Gamma_{e^+e^-} = \frac{\alpha_{\mathrm{em}}\overline{\varepsilon}^2 m_X^{}}{3}\sqrt{1-\left(\frac{2m_e}{m_X}\right)^2}\left(1+\frac{2m_e^2}{m_X^2}\right).
\end{equation}
If $X\rightarrow \gamma\gamma\gamma$ process does not deplete the dark gauge boson before $m_X$ reaches the threshold $2m_e$, then the remaining dark gauge bosons can decay into $e^+e^-$ pairs. This is one of the unique features of our model. Since the three-photon decay is strongest near $e^+e^-$ threshold [Eq.~\eqref{eq:3photondecayrate}], we expect most of the photons from $X\rightarrow \gamma\gamma\gamma$ are produced right before the beginning of the generation of $e^+e^-$ signal.

\begin{figure}[t]
    \begin{subfigure}{.5\textwidth}
    \centering\includegraphics[width=\linewidth]{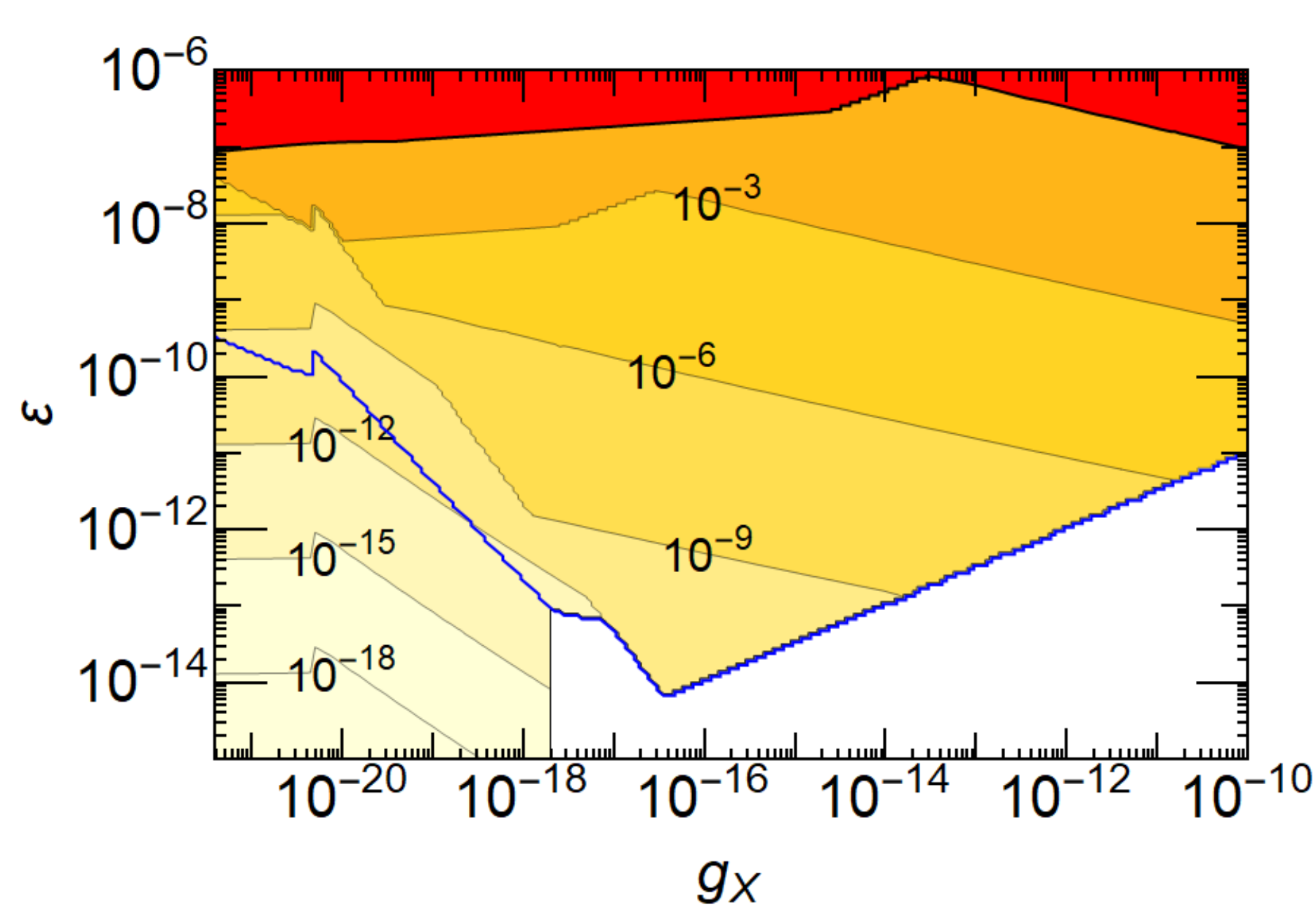}   \caption{$Y_X$ at $T=T_\text{eq}$}\label{fig:neq}\end{subfigure}
    \begin{subfigure}{.5\textwidth}\centering
    \includegraphics[width=\linewidth]{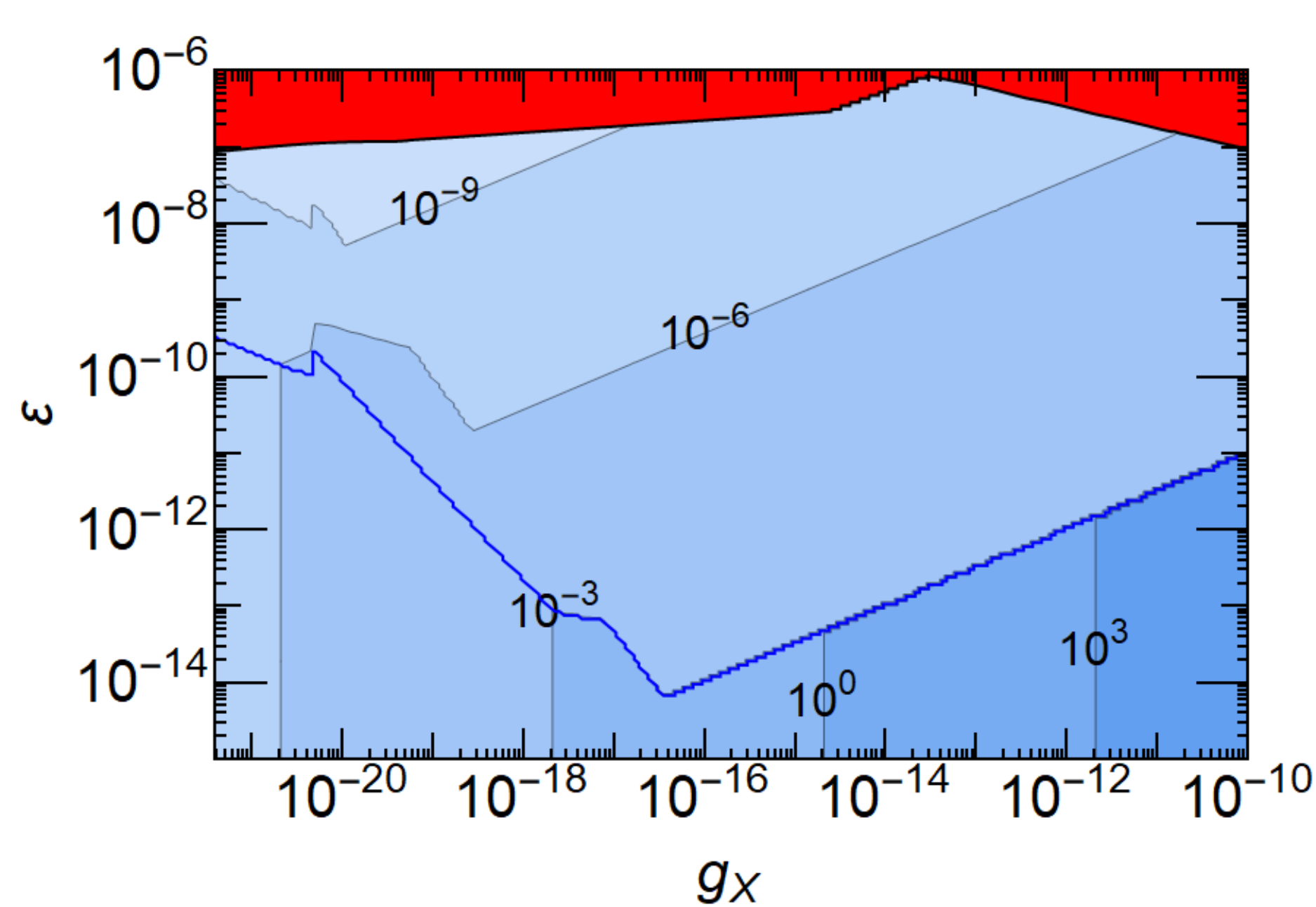} 
    \caption{$m_X$ [GeV] at $T=T_\text{eq}$}\label{fig:meq}
      \end{subfigure}
    \begin{subfigure}{.5\textwidth}
    \centering\includegraphics[width=\linewidth]{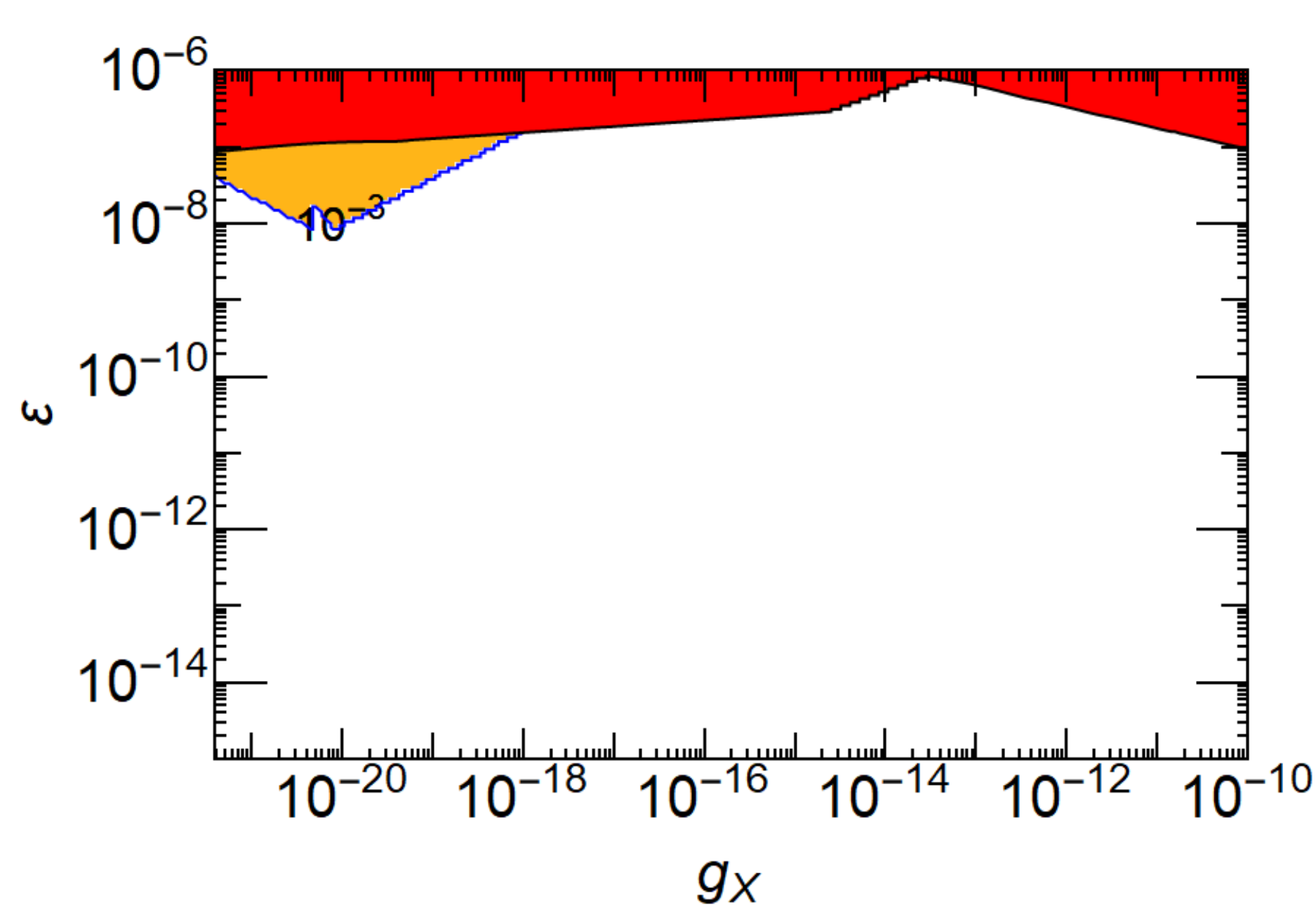} \caption{$Y_X$ at $T=T_0$}\label{fig:nnow}\end{subfigure}
    \begin{subfigure}{.5\textwidth}\centering\includegraphics[width=\linewidth]{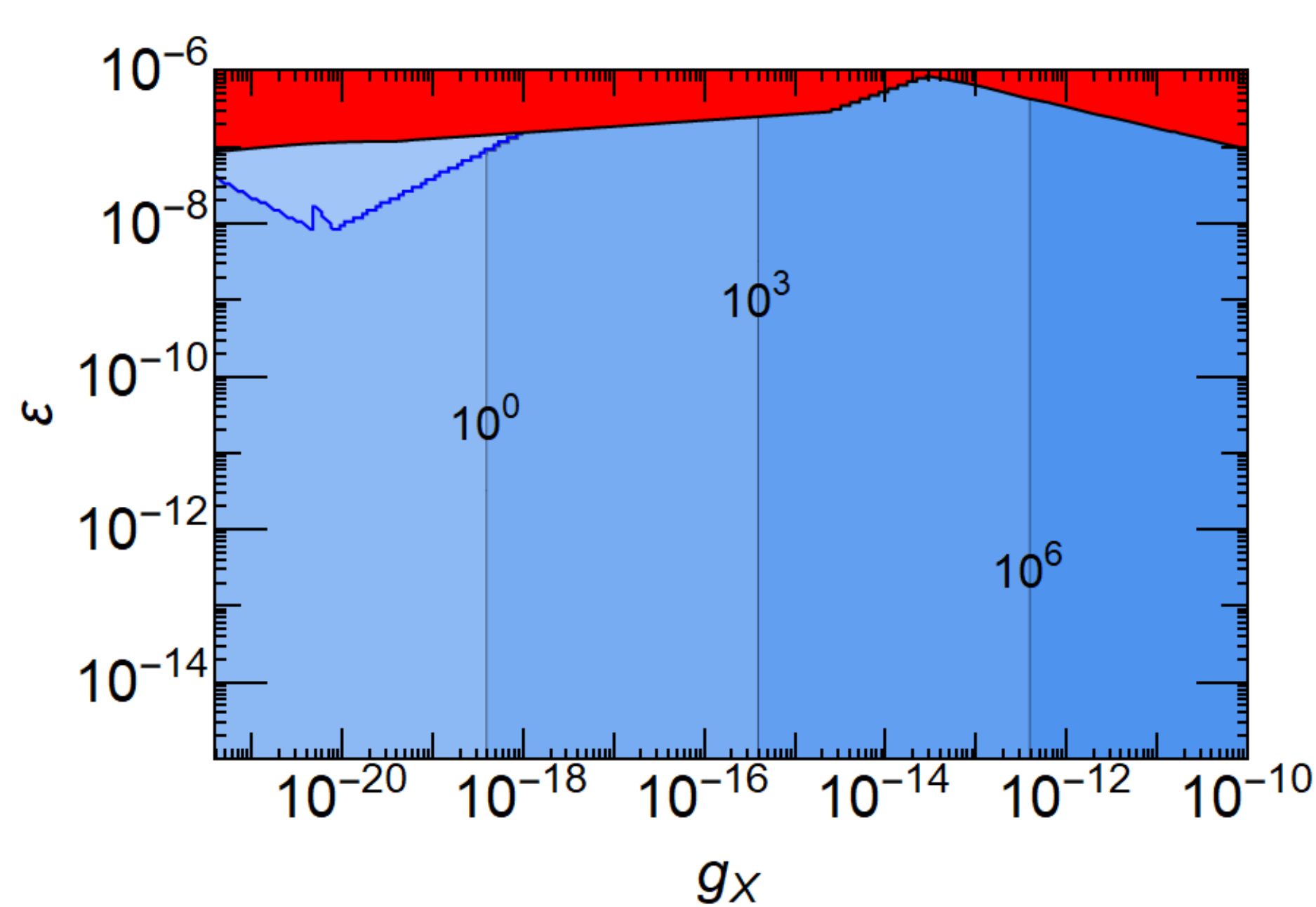} 
    \caption{$m_X$ [GeV] at $T=T_0$}\label{fig:mnow}
    \end{subfigure}    
    \caption{Time evolution of $n_X/s$ and $m_X$. Each figure corresponds to $n_X/s$ or $m_X$  at (a, b): $T=T_\text{eq}$, (c, d): $T=T_0$. 
    The blue curve corresponds to the boundary between tracking (below) and trapped (above) cases at each moment. The blank region indicates that $X$ has decayed into other components before reaching the specified time.}
\label{fig:evol}
\end{figure}

The time evolution of $Y_X$ and $m_X$ are described in Fig.~\ref{fig:evol}. As time goes by, the decay rate becomes larger than the Hubble parameter, and the decay of the dark gauge boson happens in some parameter space. The parameter space where decays happen before $T=T_\text{eq}$ and $T=T_0$ is described as the blank region in Fig.~\ref{fig:neq} and Fig.~\ref{fig:nnow}. If $X$ decays into other components, then $n_X$ becomes negligible, and $m_X$ follows the tracking solution so the contour lines below the blue curve in Fig.~\ref{fig:meq} and Fig.~\ref{fig:mnow} become vertical.

In the parameter space beyond the left border of Fig.~\ref{fig:evol}, $X$ can remain as a remnant until the present. In this case, the dark gauge bosons exist as mass-varying dark matter in the universe.

\section{Constraints}

Various studies have investigated the constraints on the $U(1)$ gauge extension of the standard model. Many of these constraints are relevant to our scenario.  However, due to the mass-varying effect, the shape of constraints is quantitatively and qualitatively different. We will discuss the constraints on our model from over-production, CMB spectrum distortion and diffuse X-ray/gamma-ray background. Figure~\ref{fig:constraint} summarizes these constraints. $\phi$ follows the tracking solution in the remaining parameter space. In this case, the present mass of dark gauge boson ($m_X^0$) is given as $3\times 10^{18}~ g^{}_X$ GeV, corresponding to the labels in the upper boundary.

\subsection{Overproduction}

If the extra energy density dominates the total energy density of the universe, it can modify the background expansion of the universe. This affects the theoretical estimation of BBN, CMB, and galaxy distributions.

\begin{figure}[tb]
    \centering
    \includegraphics[width=0.95\textwidth]{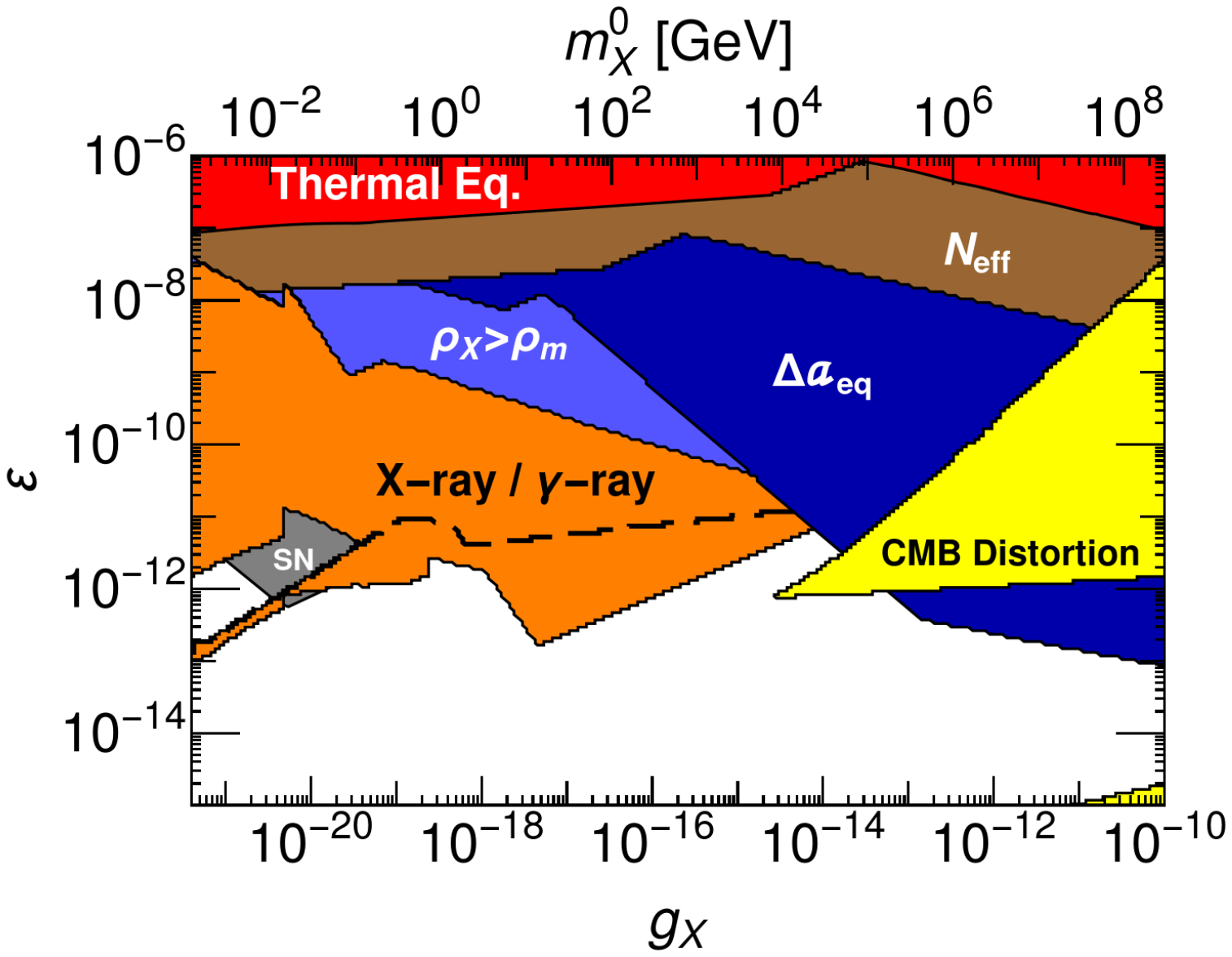}
    \caption{Constraints on $\varepsilon$ and $g_X$ when the dark gauge boson mass varies over time. The present mass of dark gauge boson ($m_X^0$) is given as $3\times 10^{18}~ g_X$ GeV. The brown region is constrained by the change of $N_{\nu}^{\mathrm{eff}}$ after the resonant production. The deep blue (light blue) region is constrained by the over-density at the matter-radiation equality (during the matter-dominated era). The gray region corresponds to the constraint from supernova \cite{Bjorken:2009mm,Chang:2016ntp,DeRocco:2019njg}. The CMB spectrum distortion constrains the yellow region, and the diffuse X-ray/gamma-ray background spectrum constrains the orange region. The small island of CMB constraint at the lower-right corner emerges due to the considerable amount of kinetic energy of $e^+$ and $e^-$ from $X$ decay, while the produced $e^+$ and $e^-$ are non-relativistic in the other region of the parameter space. Above the dashed line, the dark gauge bosons are depleted before their mass crosses the $e^+e^-$ threshold. Although we do not represent in this plot, the weak gravity conjecture \cite{Arkani-Hamed:2006emk} disfavors too tiny gauge coupling as $g_X \lesssim m_{\phi}/M_{\mathrm{Pl}} \sim 10^{-61}$. (Also, the inflationary scenario might be restricted \cite{Heidenreich:2016aqi,Heidenreich:2017sim}.) A kinetic mixing might originate from the fields that are charged under both the SM $U(1)$ gauge symmetry and $U(1)_{\mathrm{dark}}$ \cite{Holdom:1985ag}. In this case, the kinetic mixing is estimated as $\varepsilon \sim 10^{-2}  g_X$. Moreover, an even smaller kinetic mixing such as $\varepsilon \sim 10^{-13}  g_X^{} $ might be generated by the graviton-mediated interaction alone \cite{Gherghetta:2019coi}.    }
    \label{fig:constraint}
\end{figure}

Since the production of the dark gauge boson mainly occurs after the BBN, our scenario is not constrained by the BBN. However, the production of the dark gauge boson reduces the energy density and number density of the photon. Since the neutrinos are unaffected by the production of the dark gauge boson, this leads to the increase of the effective number of the neutrino ($N_{\nu}^{\mathrm{eff}}$) as \cite{Jaeckel:2008fi}
\begin{equation}
 N_{\nu}^{\mathrm{eff}} \equiv \frac{\rho_{\mathrm{rad}} -\rho_{\gamma}}{\rho_{\nu}^{\mathrm{SM}}} = \frac{3.046}{1-x}, 
\end{equation}
where $0\leq x < 1 $ is the fraction of the dark gauge boson density to the photon density right after the resonant production; $\rho_{\mathrm{rad}}$, $\rho_{\gamma}$, and $\rho_{\nu}^{\mathrm{SM}}$ is the energy density of the total radiation degree of freedom, photon, and neutrino, respectively.\footnote{If the dark gauge boson is relativistic around the CMB decoupling, there is a dark gauge boson contribution to $N_{\nu}^{\mathrm{eff}}$ \cite{Jaeckel:2008fi}. However, this contribution is negligible unless $x>\mathcal{O}(0.1)$. Since the constraint obtained from $N_{\nu}^{\mathrm{eff}}$ by considering the sole neutrino contribution is much stronger, one can ignore this contribution.} $Planck$ TT,TE,EE+lowE gives a constraint on this quantity as $N_{\nu}^{\mathrm{eff}} = 2.92^{+0.36}_{-0.37}$ \cite{Planck:2018vyg}. We give a constraint from $N_{\nu}^{\mathrm{eff}}$ by the brown region in Fig.~\ref{fig:constraint}.

An additional radiation or matter density can change the time of the matter-radiation equality. $Planck$ \cite{Planck:2018vyg} also constrains the redshift at the matter-radiation equality as $z_{\mathrm{eq}}=3402 \pm 26$. In the absence of the additional density, the scale factor\footnote{The scale factor ($a$) and the redshift ($z$) are related as $a=1/(z+1)$.} at the equality ($\bar{a}_\text{eq}$) is
\begin{equation}
\rho_\text{m}^0 {\bar{a}_\text{eq}}^{-3} = \rho_\text{rad}^0{\bar{a}_\text{eq}}^{-4},
\end{equation}
where $\rho_\text{m}^0$ ($\rho_\text{rad}^0$) is the present energy density of matter (radiation).
Then, the matter-radiation equality is shifted by the dark gauge boson density as
\begin{equation}
\rho_\text{m}^0 {a_\text{eq}}^{-3} \pm \rho_X (a_\text{eq}) = \rho_\text{rad}^0 {a_\text{eq}}^{-4},
\end{equation}
where the plus (minus) sign is applied when $X$ is non-relativistic (relativistic), respectively, and $\rho_X(a_\text{eq})$ is the energy density of $X$ at $a_\text{eq}$. Then we can find the following relation from $2$ $\sigma $ uncertainty of $z_{\mathrm{eq}}$ as
\begin{equation}
\bigg| \frac{\Delta a_\text{eq}}{a_\text{eq}} \bigg| \approx \frac{\rho_X(\bar{a}_\text{eq})}{\rho_m^0 {\bar{a}_\text{eq}}^{-3}} <\frac{52}{3403},
\end{equation}
where $\Delta a_\text{eq} =  a_\text{eq}-\bar{a}_\text{eq} $ is the shift of the matter-radiation equality. 
The constraint from the equality is described by the deep blue region in Fig.~\ref{fig:constraint}.

After the dark gauge boson becomes non-relativistic, some fraction of the dark gauge boson density over the matter density would grow due to the increasing mass. If the energy density of the dark gauge boson is comparable to or surpasses the matter density, it affects the growth of the matter density fluctuation and the lensing of the CMB photons. A detailed analysis is required to verify the actual constraints, and we will leave it to future works. Instead, we mark the light blue region in Fig.~\ref{fig:constraint}, where $\rho_{X}/\rho_m >1$ is achieved after the matter-radiation equality.\footnote{If the resonant production occurs after $T \sim 100$ eV, there might be a distortion from the resonance. However, for the parameter space of interest, the resonant production occurs when $T\gg 100$ eV, and we can neglect the distortion from the resonance. }

\subsection{CMB spectrum distortion}
\label{sec:cmbdist}
In the early universe, CMB photons obtain the blackbody spectrum from processes such as Compton scattering, double Compton scattering, and Bremsstrahlung. However, these processes become inefficient as the universe cools down. For instance, if the temperature of the SM thermal bath drops below $T \sim 100$ $\mathrm{eV}$, double Compton scattering and Bremsstrahlung processes are not effective in producing high energy photons, which leads to the loss of the chemical equilibrium. Also, the kinetic equilibrium is broken when the Compton scattering becomes inefficient from $T\sim 10$ $\mathrm{eV}$. Therefore, the energy injection to the SM thermal bath below $T \sim 100$ $\mathrm{eV}$ can cause the distortion of the CMB spectrum \cite{Hu:1992dc}.

Such a CMB spectrum distortion is highly constrained by COBE/FIRAS \cite{Fixsen:1996nj} as $|\mu| < 4.7 \times 10^{-5}$ and $|y|< 1.5 \times 10^{-5}$ \cite{Bianchini:2022dqh}. $\mu$-type distortion is mostly generated by the energy injection between the decoupling of the double Compton scattering and the Compton scattering, while $y$-type distortion is generated after the decoupling of the Compton scattering. 

When $m_X \leq 2m_e $, the dark gauge boson deposits its energy and entropy density by the three-photon decay process. In this case, $\mu$ and $y$-type distortions from the Green's function formalism are given by \cite{Chluba:2013vsa,Chluba:2013pya,Chluba:2015hma,Chluba:2016bvg}
\begin{equation}
    \mu\approx  1.401 \int \frac{\mathcal{J}_{bb}(T)\left(1-\mathcal{J}_y(T) \right)}{\rho_{\gamma}(T)} \left(\frac{1}{\rho_{\gamma}(T)}\frac{d\rho(T)}{dT}-\frac{4}{3n_{\gamma}(T)}\frac{dn(T)}{dT}\right) dT,
    \label{eq:muparam}
\end{equation}
\begin{equation}
    y \approx  \frac{1}{4} \int \frac{\mathcal{J}_y(T)}{\rho_{\gamma}(T)} \frac{d\rho(T)}{dT} dT ,
    \label{eq:yparam}
\end{equation}
where $d\rho(T)/dT$ is the power injected to the SM thermal bath, $dn(T)/dT$ is the rate of the photon entropy (number density) injection, $\rho_{\gamma}(T)$, $n_{\gamma}(T)$ are the energy density, number density of CMB photons,  and $\mathcal{J}_{bb}(T)$, $\mathcal{J}_{y}(T)$ are numerically evaluated Green's functions given by \cite{Chluba:2016bvg}
\begin{equation}
    \mathcal{J}_{bb}(z) = \exp \left[-\left(\frac{T}{T_{\mu}} \right)^{5/2}\right],
\end{equation}
\begin{equation}
    \mathcal{J}_{y}(z)= \begin{dcases}
         \left( 1+ \left(\frac{T}{T_{y}}\right)^{2.58} \right)^{-1} \quad & \mathrm{for}\; T > T_{\mathrm{eq}}, \\0 \quad &\mathrm{for} \; T<T_{\mathrm{eq}},
         \end{dcases}
\end{equation}
where $T_{\mu}=460$ eV and $T_{y}=13.9$ eV. Also, the injected power can be calculated from the following equation,
\begin{equation}\frac{d\rho}{dt} \sim \Gamma_{\gamma\gamma\gamma}m_X^{}(T)n_X(T),
\label{eq:drhodT}
\end{equation}
 and $d\rho/dt =T H(T) d\rho/dT$ from the chain rule.
$n_X(T)$ can be calculated from the Boltzmann equation, so the injected power can be written as 
\begin{equation}\frac{d\rho}{dt} \sim \Gamma_{\gamma\gamma\gamma}m_X^{}(T) \exp\left[-\frac{\Gamma_{\gamma\gamma\gamma}}{\eta H}\right]  \left(\frac{2\pi^2}{45}g_{\ast s} T^3\right) Y_X,
\label{eq:eq:drhodTFull}
\end{equation}
where $g_{\ast s}\approx 3.9$ is the effective degrees of freedom of SM thermal bath entropy density. The exponential factor describes the change of $n_X$ due to the decay, and $\eta$ is the correction due to the mass-varying nature of the dark gauge boson. (See App.~\ref{sec:Derivation of the injected power} for the detail.) 

If the dark gauge boson is not depleted before its mass reaches $2m_e$, the remaining dark gauge bosons decay into $e^+e^-$ pairs. Since electrons and baryons maintain full thermal contact with photons until $T\sim 50$ meV, the
thermal energy (kinetic energy) of $e^+$ and $e^-$ quickly kinetically equilibriate, leading to the CMB distortion through the energy injection term in Eqs.~\eqref{eq:muparam} and \eqref{eq:yparam}. As we will demonstrate in Sec.~\ref{sec:Dark energy signal}, $X\rightarrow e^+e^-$ decay occurs in a short period near $e^+e^-$ threshold. Thus, we approximate the decays occur mostly at $T_{\mathrm{dec}}$, where $\Gamma_{e^+e^-}(T_{\mathrm{dec}}) = H(T_{\mathrm{dec}})$, and the kinetic energy of a single $e^+$ or $e^-$ is given by
\begin{equation}
    E_k \sim \frac{1}{2m_e} \left(\frac{1}{4}m_X^2(T_{\mathrm{dec}}) -m_e^2\right).
\end{equation}
Consequently, the energy injected into the CMB photons is expressed as
\begin{equation}
     \Delta \rho \sim \Delta n_{e^+e^-}(E_k - \frac{3}{2} T_{\mathrm{dec}}),
\end{equation}
where $\Delta n_{e^+e^-}$ represents the number density of $e^+$ and $e^-$ released by $X\rightarrow e^+e^-$ decays.

The constraint from the CMB spectrum distortion is represented by the yellow region in Fig.~\ref{fig:constraint}. Since $e^+$ and $e^-$ generated from the dark gauge boson decay are non-relativistic, this constraint mostly arises from the $X\rightarrow \gamma\gamma\gamma$ processes. One exception arises at the lower-right corner of Fig.~\ref{fig:constraint}. In this regime, $e^+e^-$ decay occurs when $m_X$ is on the order of a few MeV, leading to the production of $e^+$ and $e^-$ with substantial kinetic energy. 
In contrast to the common idea that the stronger coupling gives a stronger constraint, the CMB spectrum distortion does not constrain the large $\varepsilon$ regime. This is because the larger $\varepsilon$ leads to the larger dark gauge boson density, which suppresses the mass of the dark gauge boson (see Fig.~\ref{fig:meq}). Thus the three-photon decay is not efficient [Eq.~\eqref{eq:3photondecayrate}] in the early universe.

\subsection{Diffuse X-ray/gamma-ray}

 The photon produced by the decay of the dark gauge boson constitutes an isotropic diffuse photon background. Such a diffuse X-ray/gamma-ray background spectrum was measured by various missions in the frequency window between $1$ keV to $10$ GeV \cite{gendreau1995asca,kappadath1998measurement,Gruber:1999yr,Strong:2004de,Bouchet:2008rp,Ajello:2008xb,Ackermann2012}. Therefore, $X\rightarrow \gamma\gamma\gamma$ signal in this frequency band can give a significant constraint.

The X-ray/gamma-ray flux from the decay of the dark gauge boson is given by \cite{Masso:1999wj,Abazajian:2001vt,Boyarsky:2005us,Bertone:2007aw,Essig:2013goa,An:2014twa}
\begin{equation}
    \frac{d^2\Phi_{\gamma}}{d\Omega dE} = \frac{3}{4\pi} \int  \frac{dN}{dE(z)} \frac{n_X(z)}{(1+z)^3} \frac{\Gamma_{\gamma\gamma\gamma}(z)}{H(z)} dz,
\end{equation}
where $dN/dE(z)$ is the spectrum of the photon from the dark gauge boson decay at the redshift $z$, and $E(z)=E_0(1+z)$ is the energy of the photon at $z$ which is detected with the energy $E_0$.  If one assumes the typical monochromatic decay\footnote{An inclusion of actual decay spectrum can alter the flux \cite{An:2014twa}. However, the one-photon spectrum is mostly distributed around $E(z) \sim m^{}_X/3$ \cite{Pospelov:2008jk}, so the monochromatic decay approximation is valid. } \cite{Yuksel:2007dr,Redondo:2008ec}, i.e. $dN/dE(z) = \delta(E(z)-m_X^{}/3)$, the simplified formula is given by 
\begin{equation}
    \frac{d^2\Phi_{\gamma}}{d\Omega dE}= \frac{3}{4\pi} \frac{\Gamma_{\gamma\gamma\gamma}(z^{\ast}) n_X(z^{\ast}
    )}{E_0 H(z^{\ast})(1+z^{\ast})^3},
    \label{eq:diffusefor}
\end{equation}
where $z^{\ast}$ is the redshift which satisfying $E(z^{\ast})=m_X^{}(z^{\ast})/3$. Unlike the conventional decaying dark matter models\footnote{We classify conventional decaying dark matter models as those where the couplings and masses of particles involved in the decay processes are fixed, and the mother particles are non-relativistic.} where the mass of the particles is fixed, the slope of the measured flux is affected by the time-variation of $\Gamma_{\gamma\gamma\gamma}$. We take that the three-photon decay terminates when $m_X^{}$ crosses $e^+e^-$ threshold or the three-photon decay rate becomes larger than the Hubble parameter. The constraint from the diffuse X-ray/gamma-ray from the collection of data in Ref.~\cite{Ajello:2008xb} is given by the orange region in Fig.~\ref{fig:constraint}. The orange region above the dashed line is where the dark gauge bosons are depleted before $m_X$ crosses the $e^+e^-$ threshold. In this regime, the overall magnitude of the signal is weaker because the decay occurs below the $e^+e^-$ threshold [Eq.~\eqref{eq:3photondecayrate}]. On the other hand, below the dashed line, the dark gauge bosons are not depleted before they reach the $e^+e^-$ threshold. Therefore, the decay mostly happens near the threshold where the decay rate is maximal. This is why the constraint appears in the small $\varepsilon$ region, while some of the larger $\varepsilon$ parameter space is not constrained.

\section{Dark energy signal}\label{sec:Dark energy signal}

So far, we have discussed how dark energy can affect the production and decay of dark gauge bosons. In this section, we will explore how the effects of dark energy can manifest in distinct non-gravitational signals.

\begin{figure}[t]
     \begin{subfigure}{0.48\textwidth}
   \centering
         \includegraphics[width=\textwidth]{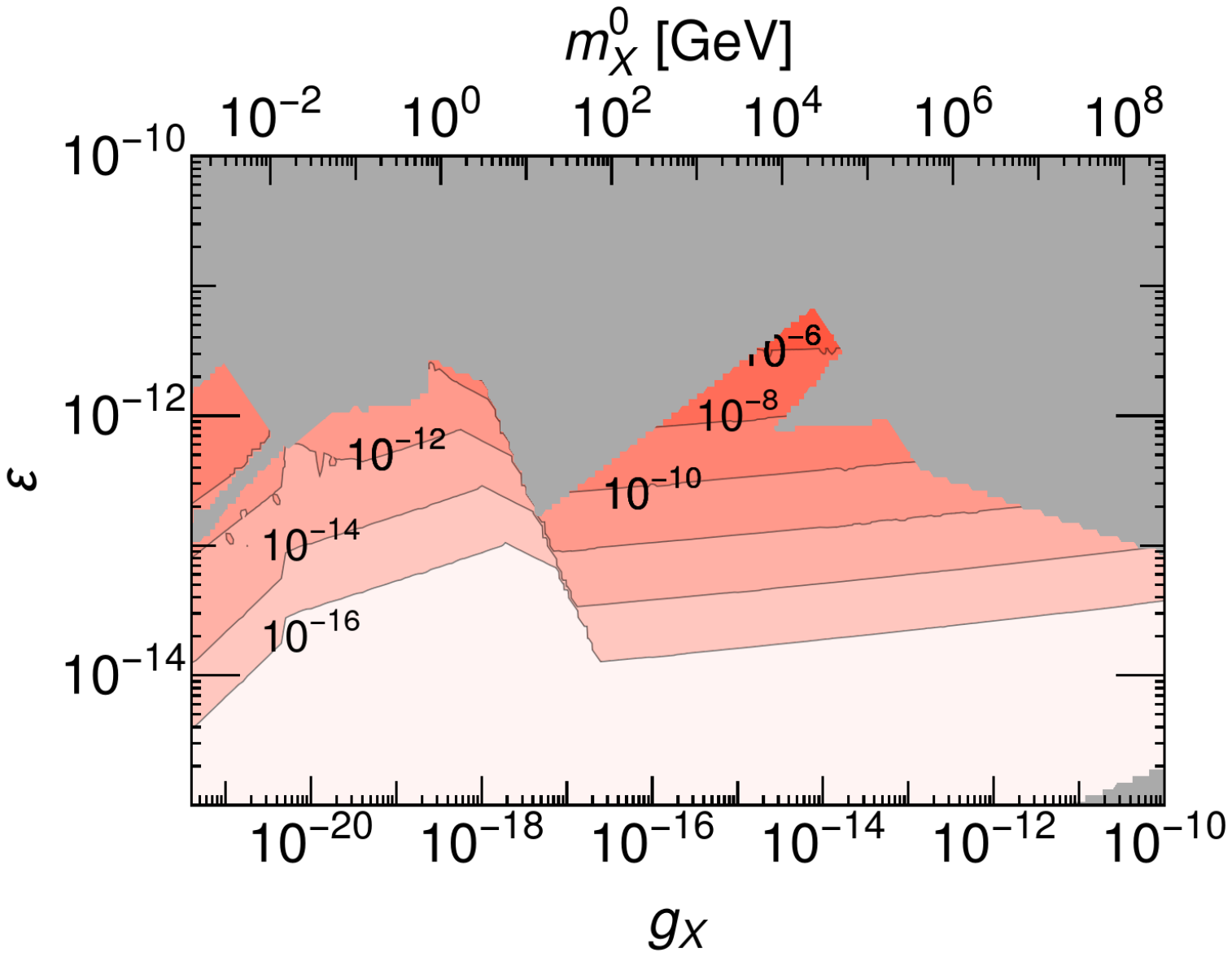}
         \caption{$\rho_{\gamma\gamma\gamma}^0/\rho_\text{crit}$}
     \end{subfigure}
   \begin{subfigure}{0.48\textwidth}
   \centering
   \includegraphics[width=\textwidth]{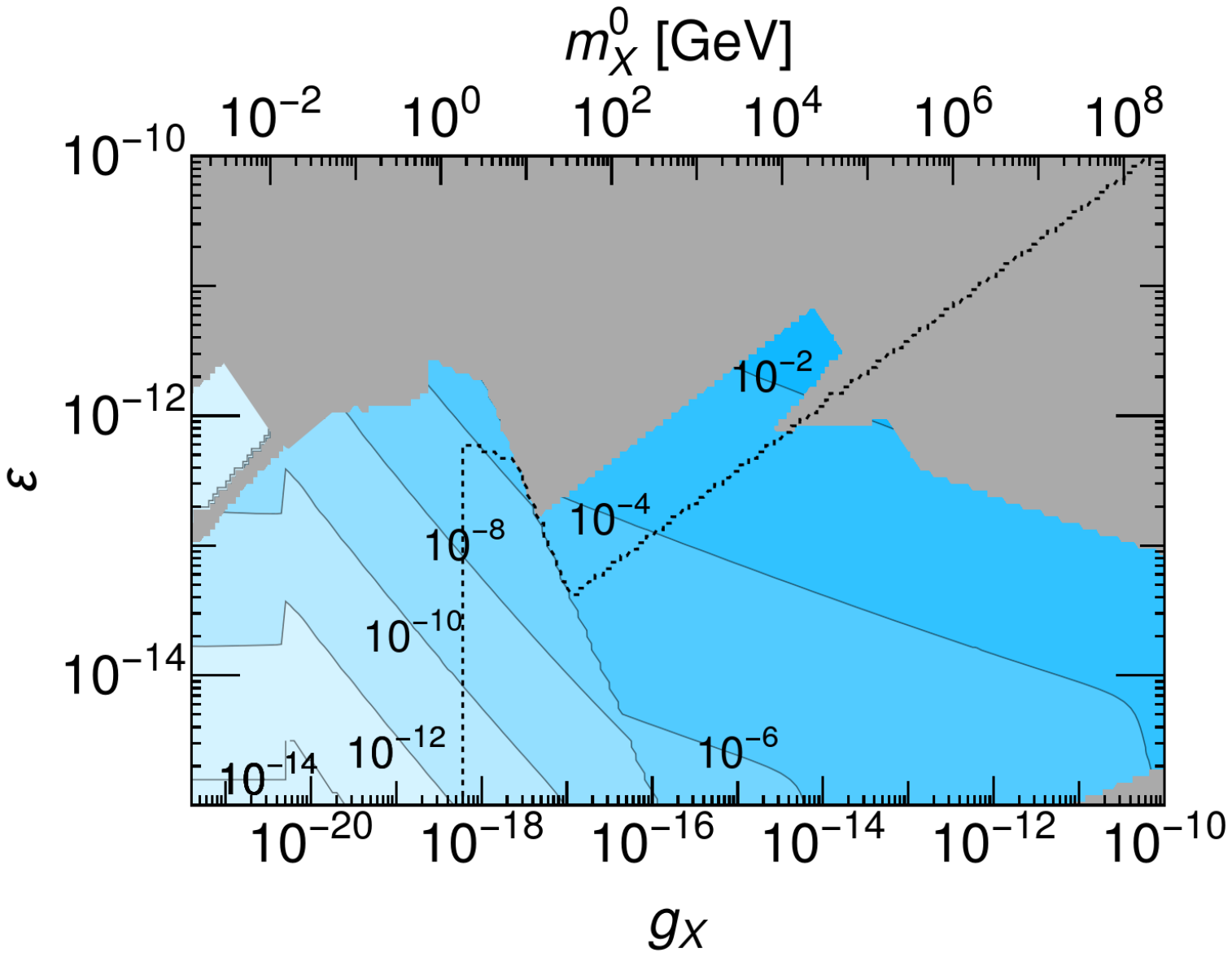}\caption{$\rho_{e^+e^-}^0/\rho_\text{crit}$}
     \end{subfigure}  
    \caption{The present energy density of the photon signal (red contours) and $e^+$/$e^-$ signal (blue contours) produced by the dark gauge boson decay. The present mass of the dark gauge boson ($m_X^0$) is given as $3\times 10^{18}~ g_X$ GeV. Both energy densities are normalized by the critical density $\rho_\text{crit}\simeq 3\times 10^{-47}$ GeV$^4$. The gray region represents the collection of the constraints in Fig.~\ref{fig:constraint}. We do not show the contours below the $\rho^0/\rho_{\mathrm{crit}}=10^{-16}$. Above the dotted curve, $e^+/e^-$ signal is produced after the recombination ($z\sim 1100$). }
    \label{fig:signal}
\end{figure}

As we mentioned, the dark gauge boson can decay into three photons or $e^+e^-$ pair. In the case of the conventional dark photon model, the decay of the gauge boson can either produce $e^+/e^-$ signal or three photon signal depending on its mass, but it cannot have both signals simultaneously \cite{Redondo:2008ec}. In our scenario, however, the increasing mass of the dark gauge boson due to the rolling of the quintessence field dark energy field allows the dark gauge boson to decay into three photons when the mass of the dark gauge boson is smaller than $2m_e$, and it decays to $e^+e^-$ pair as the mass of dark gauge boson becomes larger than $2m_e$. Therefore, both signals can be produced together. This is demonstrated in Fig~\ref{fig:signal} which shows the present energy density of decay products $\rho_{\gamma\gamma\gamma}^0$ and $\rho_{e^+e^-}^0$ normalized by the critical density $\rho_\text{crit}\simeq 3\times 10^{-47}$ GeV$^4$. The left border of the Fig.~\ref{fig:signal} represents the kinematic threshold, beyond which the production of $e^+/e^-$ signal is not possible, because the mass of the dark gauge boson is below the kinematic threshold. Except for the small region on the left corner surrounded by the gray constraint, both $e^+/e^-$ and three photon signal can be produced together.\footnote{In this region, the dark gauge bosons deplete before they cross the $e^+e^-$ threshold.}

The existing constraint from diffuse X-ray/gamma-ray overlaps with the region where $e^+$/$e^-$ density could be produced. This suggests that future technical advancements may enable the exploration of photonic channels toward the unconstrained parameter space. If $e^+$/$e^-$ signals can be probed as well, it would lead to an interesting discovery. Cosmic ray $e^+$/$e^-$ spectrum has been widely observed by various missions such as PAMELA \cite{PAMELA:2008gwm}, Fermi-LAT \cite{Fermi-LAT:2011baq}, AMS \cite{AMS:2013fma}, HESS \cite{HESS:2017clw}, DAMPE \cite{DAMPE:2017fbg}, and CALET \cite{CALET:2017uxd}. Although our $e^+$/$e^-$ signals are highly non-relativistic, they could be accelerated by the strong magnetic field \cite{Hillas1984} and contribute as an excess $e^+$/$e^-$ signal. Also, they might leave observable imprints through the scattering with baryons or high energy cosmic rays. We also note that $e^+$$/
  e^-$ signal can be generated after the recombination, above the dotted line in Fig.~\ref{fig:signal},
while conventional dark photon model with $\varepsilon > 
 10^{-15} $ cannot produce the background $e^+$$/
   e^-$ signal after the recombination ($z\sim 1100$).  

\begin{figure}[t]
   \centering
    \includegraphics[width=0.8\textwidth]{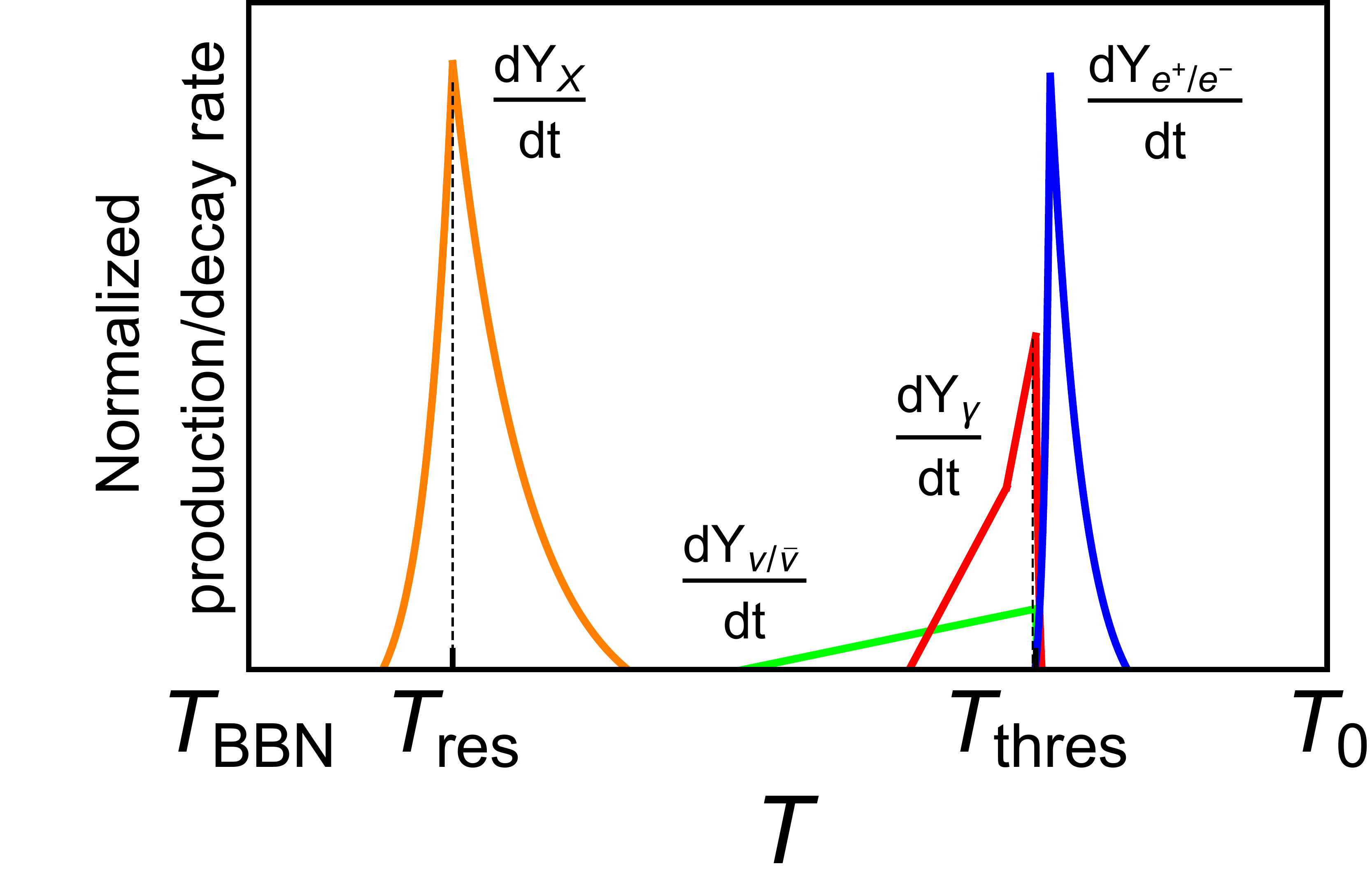}
    \caption{A schematic description of the production and decay of the dark gauge boson. $Y_i$ denotes the normalized number density of each particle species $i=\{X,\nu,\gamma,e\}$. The orange curve is the normalized production rate of the dark gauge boson: $dY_X/dt$, and each of red, blue, green curve is the production rate of the photon: $dY_{\gamma}/dt$, positron/electron: $dY_{e^+/e^-}/dt$, neutrino/anti-neutrino: $dY_{\nu/\bar{\nu}}/dt$ from the decay of the dark gauge boson. The dark gauge bosons are resonantly produced at $T_{\mathrm{res}}$, and productions of the photon, neutrino and $e^+/e^-$ signals mostly occur near $e^+e^-$ threshold $T_{\mathrm{thres}}$ when $m_X \sim 2m_e$ is achieved.}  
    \label{fig:sch}
\end{figure}

Figure~\ref{fig:sch} is the schematic description of the production and decay of the dark gauge boson. Normalized production/decay rate of each species ($dY_i/dt$, where $Y_i$ is the normalized energy density of each species $i=\{X,\nu,\gamma,e\}$) shows the chronological order of each event. Since the three-photon decay rate has a steep dependence on $m_X$ ($\Gamma_{\gamma\gamma\gamma}\propto m_X^9$), most of the three-photon decay might happen near $e^+e^-$ threshold. The produced energy density of the photon signal is calculated by solving the Boltzmann equation. If $m_X^{}$ crosses the threshold before the dark gauge bosons are depleted, the remaining dark gauge bosons can decay into $e^+e^-$ pairs. Interestingly, we found that the remaining dark gauge bosons almost instantaneously decay to $e^+e^-$ pairs, i.e. $\Gamma_{e^+e^-} > H$, near $e^+e^-$ threshold. This is because the $X\rightarrow e^+e^-$ process is blocked, resembling water contained behind a sealed dam, until the kinematic threshold is surpassed.\footnote{An exception appears only at the small portion of the bottom right corner of Fig.~\ref{fig:signal}. However, even in this region, the decay occurs when $m_X$ and the energy of resultant $e^+e^-$ are a few MeV.} Therefore, the remaining dark gauge boson energy density is quickly converted to the $e^+e^-$ pair energy density. Also, the produced $e^+$ and $e^-$ are non-relativistic since they are produced right at the threshold. It could be the signal from the gauged dark energy model if one can observe the exotic photonic and $e^+$/$e^-$ signal together.

\begin{figure}[t]
    \centering
    \includegraphics[width=0.8\textwidth]{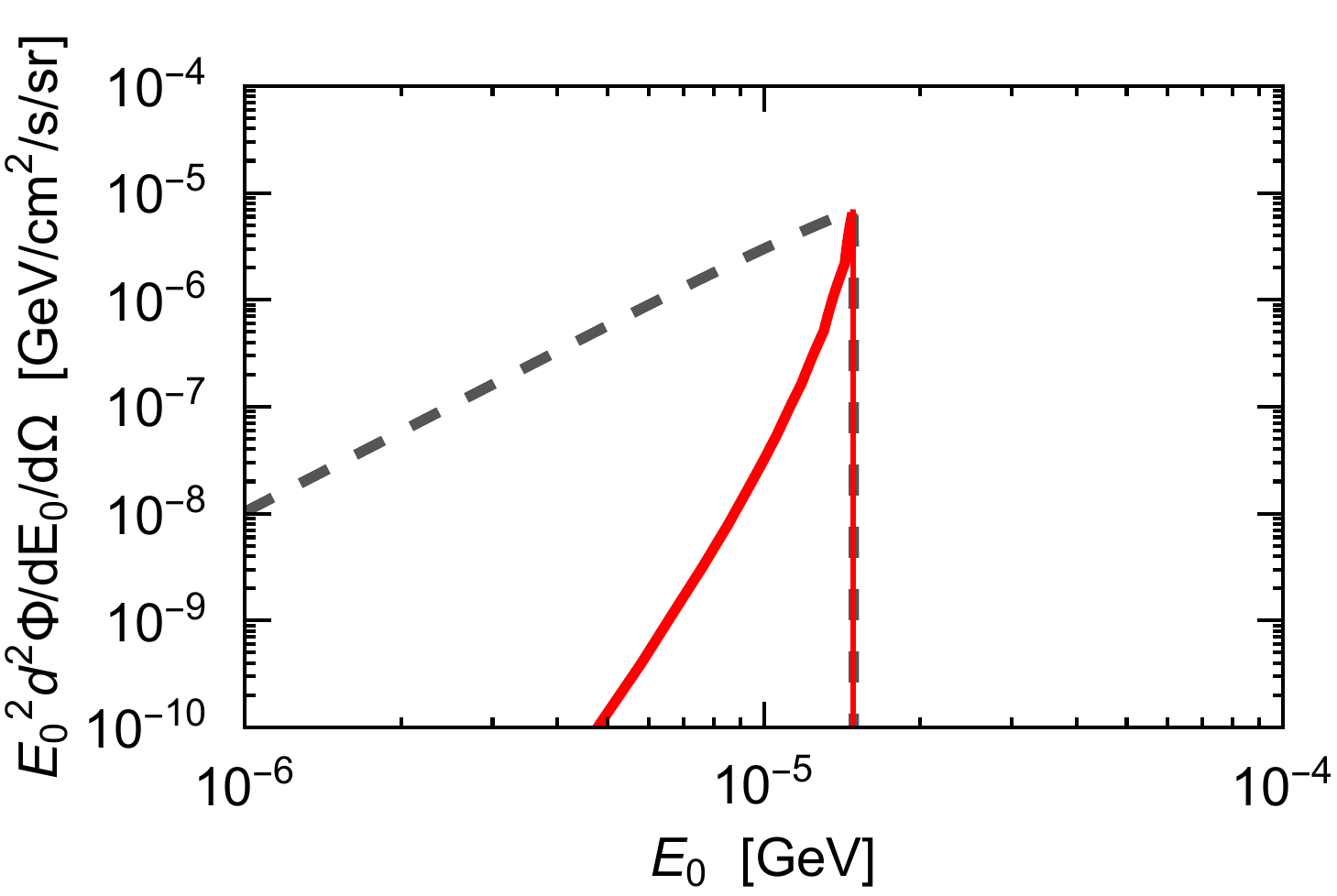}
    \caption{The diffuse X-ray/gamma-ray spectrum from the gauged quintessence (red) and conventional decaying dark matter model (dashed gray). The red curve represents the actual spectrum with $\alpha=1$, $g_X^{}=10^{-20}$, $\varepsilon= 6.3\times 10^{-13}$, while the gray curve is rescaled to give the same peak energy and peak signal magnitude to the red curve. For any conventional decaying dark matter model, regardless of the mass of the decaying particle or the decay rate, the slope of the gray curve is proportional to $(E/E_{\mathrm{peak}})^{5/2}/\sqrt{1+3(E/E_{\mathrm{peak}})^3}$, where $E_{\mathrm{peak}}$ is the peak energy \cite{Essig:2013goa}. The red curve is steeper because $\Gamma_{\gamma\gamma\gamma}$ has a sharp dependence on $m_X^{}$.}
    \label{fig:DiffSpec}
\end{figure}

Also, the gauged quintessence model can have a unique diffuse X-ray/gamma-ray spectrum.  Figure~\ref{fig:DiffSpec} shows the spectrum of the photon arising from the decaying particles both from the gauged quintessence model and the conventional decaying dark matter models. In the conventional model, the overall X-ray/gamma-ray signal from $n$-body process is proportional to the decay rate ($\Gamma_{\mathrm{dm}}$), number density of the dark matter particle ($n_{\mathrm{dm}}$), and mass of the dark matter ($m_{\mathrm{dm}}$). If one assumes the monochromatic decay, then the decay signal is peaked at $m_{\mathrm{dm}}/n$, and the measured flux is proportional to $\sqrt{E/E_{\mathrm{peak}}}/\sqrt{1+3(E/E_{\mathrm{peak}})^3}$. Therefore, the slope of the measured X-ray/gamma-ray spectrum is universal regardless of the detail of the model.

However, the diffuse X-ray/gamma-ray spectrum in the gauged quintessence model could be completely different from the conventional model. Since the decay rate is proportional to very large powers of $m_X^{}$ ($\Gamma_{\gamma\gamma\gamma} \propto m_X^{9}$), the decay rate rapidly grows with increasing $m_X^{}$. This results in the higher energy part of the diffuse photon spectrum, generated in a more recent era, being amplified compared to the lower energy part of the spectrum. Therefore, the spectrum's slope becomes much steeper than the conventional decaying dark matter model.

Before closing this section, we briefly comment on the neutrino signal in our scenario. Neutrinos can also be produced by $X\rightarrow \nu\bar{\nu}$ process. Generated neutrino flux can be calculated from Eq.~\eqref{eq:diffusefor} by replacing $3\Gamma_{\gamma\gamma\gamma}$ to $2\Gamma_{\nu\overline{\nu}}$.  Similar to the three-photon decay rate,  $\Gamma_{\nu\bar{\nu}}$ also grows with increasing $m_X$ ($\Gamma_{\nu\bar{\nu}}\propto m_X^5$). Therefore, the diffuse neutrino spectrum also becomes sharper by the mass-varying effect. Although the neutrino decay spectrum is too small to be detected by current observations,\footnote{The neutrino decay rate may be enhanced if the dark gauge boson has an additional mass mixing \cite{Davoudiasl:2012ag,Davoudiasl:2014kua,Davoudiasl:2023cnc}.} they remain as the potentially interesting observational window.

\section{Summary and conclusion}\label{Sec:summary}

In this work, we investigated non-gravitational signals from the gauged dark energy sector with a realization in the gauged quintessence model. The dark sector (quintessence field scalar + dark gauge boson) is connected to the SM sector via portals, and we studied the vector portal case. The time-varying dark energy field makes the mass of the dark gauge boson evolve during cosmic history. Since the production and decay of the dark gauge boson are deeply related to its mass, this makes the phenomenology of the dark gauge boson completely different from the conventional dark photon model of the same vector portal. For instance, the large number density of the dark gauge boson can backreact to the dynamics of the quintessence field dark energy. This makes the resonance production of the dark gauge boson in the thermal plasma enhanced through the extended resonance stage.

One of the interesting features of our scenario is that the decay channel of the dark gauge boson to the electron-positron pair might be closed in the early universe because the mass of the dark gauge boson was below $e^+e^-$ kinematic threshold. So, the dark gauge boson decays to the three-photon and neutrinos in the early universe, but the remaining dark photon is converted to $e^+e^-$ pair when the mass of the dark gauge boson crosses $e^+e^-$ threshold. Therefore, we may have the exotic relic photons, neutrinos, and $e^+e^-$ together. If the exotic relic photons/neutrinos produced by the decay of the dark gauge boson contribute to the diffuse photon/neutrino background, its spectrum could be distinguished from the conventional decaying dark matter signal. We also expect $e^+$/$e^-$ might give the excess cosmic $e^+$/$e^-$ flux, or produce secondary signals by the scattering with baryons or high energy cosmic rays.

So far, the research on dark energy phenomenology has mainly focused on the gravitational effect of dark energy through the modification of the universe's expansion near the present. Through this work, we emphasize that dark energy can significantly affect other sectors of the universe by non-gravitational interaction based on the gauge principle, and we might be able to explore the physics of dark energy with those signals.

\begin{acknowledgments}
We thank Daniel Chung for the helpful discussions.
This work was supported in part by the JSPS KAKENHI (Grant No. 20H00160) and the National Research Foundation of Korea (Grant No. NRF-2021R1A2C2009718).
\end{acknowledgments}

\appendix

\section{Diagonalization of kinetic and mass terms}\label{sec:Diagonalization of kinetic and mass terms}

The kinetic terms for the Abelian gauge fields can be diagonalized by the following rotation of the fields as
\begin{equation}
    \begin{bmatrix}
    \hat{B}_{\mu}\\ \hat{X}_{\mu}
     \end{bmatrix}  =    \begin{bmatrix}
     1 & \eta \frac{\varepsilon}{c_W}\\0& \eta
     \end{bmatrix}     \begin{bmatrix}
    \tilde{B}_{\mu}\\ \tilde{X}_{\mu}
     \end{bmatrix},
\end{equation}
where $c_W\equiv \cos\theta_W$, $\eta=1/\sqrt{1-\varepsilon^2/c_W^2}$. Then, the kinetic terms for the Abelian gauge fields are diagonal as
\begin{equation}
    \mathcal{L}_{\mathrm{kin}} \supset -\frac{1}{4}\tilde{B}_{\mu\nu}\tilde{B}^{\mu\nu}-\frac{1}{4}\tilde{X}_{\mu\nu}\tilde{X}^{\mu\nu}.
\end{equation}
However, the mass terms for $Z$ boson and $X$ boson are not diagonal on this basis as
\begin{equation}
\mathcal{L}_{\mathrm{mass}} \supset  \frac{\hat{m}_Z^2}{2}
    \begin{bmatrix} 
    \tilde{Z}_{\mu} & \tilde{X}_{\mu}
    \end{bmatrix} 
    \begin{bmatrix}
        1 & -\Delta \\ -\Delta & \eta^2r_0+\Delta^2
    \end{bmatrix}
     \begin{bmatrix} 
    \tilde{Z}_{\mu} \\ \tilde{X}_{\mu}
    \end{bmatrix}, 
    \label{eq:massMatrix}
\end{equation}
where $\hat{m}_Z^2=v^2\sqrt{g^2+g^{\prime2}}/4$ with $g$, $g^{\prime}$ being the gauge coupling of $SU(2)_{\text{L}}$, $U(1)_{\text{Y}}$, respectively, $\Delta\equiv \eta\varepsilon t_W$, and $r_0=\hat{m}^2_X/\hat{m}^2_Z$.

The mass matrix can be diagonalized by an additional rotation matrix,
\begin{equation}
        \begin{bmatrix}
  Z_{\mu}\\ X_{\mu}
    \end{bmatrix}  =    \begin{bmatrix}
     \cos\theta_a & - \sin\theta_a\\\sin\theta_a & \cos\theta_a
     \end{bmatrix}     \begin{bmatrix}
    \tilde{Z}_{\mu}\\ \tilde{X}_{\mu}
     \end{bmatrix},
\end{equation}
where $\theta_a$ is determined as 
\begin{equation}
\begin{split}
    \tan2\theta_a &= \frac{2\Delta}{1-\eta^2r_0-\Delta^2},\\
    \sin\theta_a&\approx \varepsilon t_W(1+r_0),\\
    \cos\theta_a&\approx 1-\varepsilon^2t_W^2(1+2r_0).
\end{split}
\end{equation}

\section{Derivation of the injected power formula}\label{sec:Derivation of the injected power}

In this section, we give a detailed derivation of Eq.~\eqref{eq:eq:drhodTFull}, which we use to calculate the energy injection into the SM thermal bath. The Boltzmann equation of $n_X$ after the resonance production can be written as
\begin{equation}
    \dot{n}_X(t) +3H(t) n_X(t) = -\Gamma_{\gamma\gamma\gamma}(t) n_X(t),
\end{equation}
where $\Gamma_{\gamma\gamma\gamma}$ has time dependence via $m_X$.

The exact solution of this differential equation is simply given by
\begin{equation}
    n_X^{}(t)=n_X^{}(t_i) \left(\frac{a(t_i)}{a(t)}\right)^3 \exp\left[-\int_{T(t_i)}^{T(t)} \frac{\Gamma_{\gamma\gamma\gamma}(T^{\prime})}{H(T^{\prime})T^{\prime}} dT^{\prime}  \right],
    \label{eq:nXexact}
\end{equation}
where $t_i$ can be arbitrary time prior to $t$, and $T(t)$ is the temperature at $t$. In this expression, $(a(t_i)/a(t))^{3}$ factor describes the dilution due to the expansion of the universe, while the exponential factor corresponds to the decay. Let us suppose that the decay rate is proportional to some negative power of the temperature, i.e. $\Gamma_{\gamma\gamma\gamma}/(HT)\propto T^{-\beta}$ with $\beta>0$. In this case, the integration can be easily carried out, and the result is only sensitive to the information at $t$, if $t$ and $t_i$ are well separated. Then Eq.~\eqref{eq:nXexact} can be written as
\begin{equation}
    n_X^{}(t) \approx \exp\left[-\frac{\Gamma_{\gamma\gamma\gamma}(t)}{(\beta-1) H(t)}\right]  \left(\frac{2\pi^2}{45}g_{\ast s} T^3 \right)Y_X.
    \label{eq:nXapprox}
\end{equation} 

If $m_X$ is in the tracking regime or the trapped regime where the exponential factor of $n_X(t)$ is close to the one, $m_X$ as well as $\Gamma_{\gamma\gamma\gamma}$ should be proportional to some negative powers of $T$. The loophole of the approximation arises when i) the exponential factor of $n_X(t)$ becomes much larger than one during the trapped regime, or ii) near $e^+e^-$ threshold where the enhancement factor $F(m_X)$ has a significant impact on the decay rate. 

Nonetheless, we can use Eq.~\eqref{eq:nXapprox} to estimate the energy injection into the SM thermal bath. In case i), the growth of the exponential factor implies that a large portion of the dark gauge boson has already decayed to the photons. Thus, the contribution of error in this regime is tiny. In case ii), the enhancement factor is important only shortly before $m_X$ passes $e^+e^-$ threshold. Since $e^+e^-$ channel opens right after the threshold, energy injection can be captured by taking that the remaining energy density decays at $e^+e^-$ threshold.

The values of $\beta$ in various situations are given by
\begin{equation}
    \beta= \begin{dcases}
         (3\alpha+30)/(\alpha+1) &(\mathrm{trapped,} \;\; T>T_{\mathrm{eq}} ),\\
         (5\alpha+59)/(2\alpha+2) &(\mathrm{trapped,} \;\; T<T_{\mathrm{eq}} ),\\
         (3\alpha+42)/(\alpha+2) &(\mathrm{tracking,} \;\; T>T_{\mathrm{eq}} ),\\
         (5\alpha+64)/(2\alpha+4) &(\mathrm{tracking,} \;\; T<T_{\mathrm{eq}} ).
    \end{dcases}
\end{equation}

\bibliographystyle{JHEP}
\bibliography{ref.bib}

\providecommand{\href}[2]{#2}\begingroup\raggedright\begin{thebibliography}{100}

\bibitem{SupernovaSearchTeam:1998fmf}
{\scshape Supernova Search Team} collaboration, \emph{{Observational evidence
  from supernovae for an accelerating universe and a cosmological constant}},
  \href{https://doi.org/10.1086/300499}{\emph{Astron. J.} {\bfseries 116}
  (1998) 1009} [\href{https://arxiv.org/abs/astro-ph/9805201}{{\ttfamily
  astro-ph/9805201}}].

\bibitem{SupernovaCosmologyProject:1998vns}
{\scshape Supernova Cosmology Project} collaboration, \emph{{Measurements of
  $\Omega$ and $\Lambda$ from 42 high redshift supernovae}},
  \href{https://doi.org/10.1086/307221}{\emph{Astrophys. J.} {\bfseries 517}
  (1999) 565} [\href{https://arxiv.org/abs/astro-ph/9812133}{{\ttfamily
  astro-ph/9812133}}].

\bibitem{Wetterich:1987fm}
C.~Wetterich, \emph{{Cosmology and the Fate of Dilatation Symmetry}},
  \href{https://doi.org/10.1016/0550-3213(88)90193-9}{\emph{Nucl. Phys. B}
  {\bfseries 302} (1988) 668}
  [\href{https://arxiv.org/abs/1711.03844}{{\ttfamily 1711.03844}}].

\bibitem{Ratra:1987rm}
B.~Ratra and P.J.E.~Peebles, \emph{{Cosmological Consequences of a Rolling
  Homogeneous Scalar Field}},
  \href{https://doi.org/10.1103/PhysRevD.37.3406}{\emph{Phys. Rev. D}
  {\bfseries 37} (1988) 3406}.

\bibitem{Ferreira:1997au}
P.G.~Ferreira and M.~Joyce, \emph{{Structure formation with a selftuning scalar
  field}}, \href{https://doi.org/10.1103/PhysRevLett.79.4740}{\emph{Phys. Rev.
  Lett.} {\bfseries 79} (1997) 4740}
  [\href{https://arxiv.org/abs/astro-ph/9707286}{{\ttfamily
  astro-ph/9707286}}].

\bibitem{Caldwell:1997ii}
R.R.~Caldwell, R.~Dave and P.J.~Steinhardt, \emph{{Cosmological imprint of an
  energy component with general equation of state}},
  \href{https://doi.org/10.1103/PhysRevLett.80.1582}{\emph{Phys. Rev. Lett.}
  {\bfseries 80} (1998) 1582}
  [\href{https://arxiv.org/abs/astro-ph/9708069}{{\ttfamily
  astro-ph/9708069}}].

\bibitem{Brandenberger:2009jq}
R.H.~Brandenberger, \emph{{Alternatives to the inflationary paradigm of
  structure formation}},
  \href{https://doi.org/10.1142/S2010194511000109}{\emph{Int. J. Mod. Phys.
  Conf. Ser.} {\bfseries 01} (2011) 67}
  [\href{https://arxiv.org/abs/0902.4731}{{\ttfamily 0902.4731}}].

\bibitem{Mukhanov:1981xt}
V.F.~Mukhanov and G.V.~Chibisov, \emph{{Quantum Fluctuations and a Nonsingular
  Universe}}, {\emph{JETP Lett.} {\bfseries 33} (1981) 532}.

\bibitem{Kodama:1984ziu}
H.~Kodama and M.~Sasaki, \emph{{Cosmological Perturbation Theory}},
  \href{https://doi.org/10.1143/PTPS.78.1}{\emph{Prog. Theor. Phys. Suppl.}
  {\bfseries 78} (1984) 1}.

\bibitem{Alpher:1948ve}
R.A.~Alpher, H.~Bethe and G.~Gamow, \emph{{The origin of chemical elements}},
  \href{https://doi.org/10.1103/PhysRev.73.803}{\emph{Phys. Rev.} {\bfseries
  73} (1948) 803}.

\bibitem{Hayashi:1950lqo}
C.~Hayashi, \emph{{Proton-Neutron Concentration Ratio in the Expanding Universe
  at the Stages preceding the Formation of the Elements}},
  \href{https://doi.org/10.1143/ptp/5.2.224}{\emph{Prog. Theor. Phys.}
  {\bfseries 5} (1950) 224}.

\bibitem{Walker:1991ap}
T.P.~Walker, G.~Steigman, D.N.~Schramm, K.A.~Olive and H.-S.~Kang,
  \emph{{Primordial nucleosynthesis redux}},
  \href{https://doi.org/10.1086/170255}{\emph{Astrophys. J.} {\bfseries 376}
  (1991) 51}.

\bibitem{Damour:1994ya}
T.~Damour and A.M.~Polyakov, \emph{{String theory and gravity}},
  \href{https://doi.org/10.1007/BF02106709}{\emph{Gen. Rel. Grav.} {\bfseries
  26} (1994) 1171} [\href{https://arxiv.org/abs/gr-qc/9411069}{{\ttfamily
  gr-qc/9411069}}].

\bibitem{Damour:1994zq}
T.~Damour and A.M.~Polyakov, \emph{{The String dilaton and a least coupling
  principle}}, \href{https://doi.org/10.1016/0550-3213(94)90143-0}{\emph{Nucl.
  Phys. B} {\bfseries 423} (1994) 532}
  [\href{https://arxiv.org/abs/hep-th/9401069}{{\ttfamily hep-th/9401069}}].

\bibitem{Gasperini:2001pc}
M.~Gasperini, F.~Piazza and G.~Veneziano, \emph{{Quintessence as a runaway
  dilaton}}, \href{https://doi.org/10.1103/PhysRevD.65.023508}{\emph{Phys. Rev.
  D} {\bfseries 65} (2002) 023508}
  [\href{https://arxiv.org/abs/gr-qc/0108016}{{\ttfamily gr-qc/0108016}}].

\bibitem{Choi:1999xn}
K.~Choi, \emph{{String or M theory axion as a quintessence}},
  \href{https://doi.org/10.1103/PhysRevD.62.043509}{\emph{Phys. Rev. D}
  {\bfseries 62} (2000) 043509}
  [\href{https://arxiv.org/abs/hep-ph/9902292}{{\ttfamily hep-ph/9902292}}].

\bibitem{Kim:2002tq}
J.E.~Kim and H.P.~Nilles, \emph{{A Quintessential axion}},
  \href{https://doi.org/10.1016/S0370-2693(02)03148-9}{\emph{Phys. Lett. B}
  {\bfseries 553} (2003) 1}
  [\href{https://arxiv.org/abs/hep-ph/0210402}{{\ttfamily hep-ph/0210402}}].

\bibitem{Jordan:1959eg}
P.~Jordan, \emph{{The present state of Dirac's cosmological hypothesis}},
  \href{https://doi.org/10.1007/BF01375155}{\emph{Z. Phys.} {\bfseries 157}
  (1959) 112}.

\bibitem{Brans:1961sx}
C.~Brans and R.H.~Dicke, \emph{{Mach's principle and a relativistic theory of
  gravitation}}, \href{https://doi.org/10.1103/PhysRev.124.925}{\emph{Phys.
  Rev.} {\bfseries 124} (1961) 925}.

\bibitem{Bergmann:1968ve}
P.G.~Bergmann, \emph{{Comments on the scalar tensor theory}},
  \href{https://doi.org/10.1007/BF00668828}{\emph{Int. J. Theor. Phys.}
  {\bfseries 1} (1968) 25}.

\bibitem{Wagoner:1970vr}
R.V.~Wagoner, \emph{{Scalar tensor theory and gravitational waves}},
  \href{https://doi.org/10.1103/PhysRevD.1.3209}{\emph{Phys. Rev. D} {\bfseries
  1} (1970) 3209}.

\bibitem{Copeland:1997et}
E.J.~Copeland, A.R.~Liddle and D.~Wands, \emph{{Exponential potentials and
  cosmological scaling solutions}},
  \href{https://doi.org/10.1103/PhysRevD.57.4686}{\emph{Phys. Rev. D}
  {\bfseries 57} (1998) 4686}
  [\href{https://arxiv.org/abs/gr-qc/9711068}{{\ttfamily gr-qc/9711068}}].

\bibitem{Bartolo:1999sq}
N.~Bartolo and M.~Pietroni, \emph{{Scalar tensor gravity and quintessence}},
  \href{https://doi.org/10.1103/PhysRevD.61.023518}{\emph{Phys. Rev. D}
  {\bfseries 61} (2000) 023518}
  [\href{https://arxiv.org/abs/hep-ph/9908521}{{\ttfamily hep-ph/9908521}}].

\bibitem{Comelli:2003cv}
D.~Comelli, M.~Pietroni and A.~Riotto, \emph{{Dark energy and dark matter}},
  \href{https://doi.org/10.1016/j.physletb.2003.05.006}{\emph{Phys. Lett. B}
  {\bfseries 571} (2003) 115}
  [\href{https://arxiv.org/abs/hep-ph/0302080}{{\ttfamily hep-ph/0302080}}].

\bibitem{Khoury:2003aq}
J.~Khoury and A.~Weltman, \emph{{Chameleon fields: Awaiting surprises for tests
  of gravity in space}},
  \href{https://doi.org/10.1103/PhysRevLett.93.171104}{\emph{Phys. Rev. Lett.}
  {\bfseries 93} (2004) 171104}
  [\href{https://arxiv.org/abs/astro-ph/0309300}{{\ttfamily
  astro-ph/0309300}}].

\bibitem{Khoury:2003rn}
J.~Khoury and A.~Weltman, \emph{{Chameleon cosmology}},
  \href{https://doi.org/10.1103/PhysRevD.69.044026}{\emph{Phys. Rev. D}
  {\bfseries 69} (2004) 044026}
  [\href{https://arxiv.org/abs/astro-ph/0309411}{{\ttfamily
  astro-ph/0309411}}].

\bibitem{Brax:2004qh}
P.~Brax, C.~van~de Bruck, A.-C.~Davis, J.~Khoury and A.~Weltman,
  \emph{{Detecting dark energy in orbit: The cosmological chameleon}},
  \href{https://doi.org/10.1103/PhysRevD.70.123518}{\emph{Phys. Rev. D}
  {\bfseries 70} (2004) 123518}
  [\href{https://arxiv.org/abs/astro-ph/0408415}{{\ttfamily
  astro-ph/0408415}}].

\bibitem{Mota:2006fz}
D.F.~Mota and D.J.~Shaw, \emph{{Evading Equivalence Principle Violations,
  Cosmological and other Experimental Constraints in Scalar Field Theories with
  a Strong Coupling to Matter}},
  \href{https://doi.org/10.1103/PhysRevD.75.063501}{\emph{Phys. Rev. D}
  {\bfseries 75} (2007) 063501}
  [\href{https://arxiv.org/abs/hep-ph/0608078}{{\ttfamily hep-ph/0608078}}].

\bibitem{Faulkner:2006ub}
T.~Faulkner, M.~Tegmark, E.F.~Bunn and Y.~Mao, \emph{{Constraining f(R) Gravity
  as a Scalar Tensor Theory}},
  \href{https://doi.org/10.1103/PhysRevD.76.063505}{\emph{Phys. Rev. D}
  {\bfseries 76} (2007) 063505}
  [\href{https://arxiv.org/abs/astro-ph/0612569}{{\ttfamily
  astro-ph/0612569}}].

\bibitem{Hinterbichler:2011ca}
K.~Hinterbichler, J.~Khoury, A.~Levy and A.~Matas, \emph{{Symmetron
  Cosmology}}, \href{https://doi.org/10.1103/PhysRevD.84.103521}{\emph{Phys.
  Rev. D} {\bfseries 84} (2011) 103521}
  [\href{https://arxiv.org/abs/1107.2112}{{\ttfamily 1107.2112}}].

\bibitem{Wang:2012kj}
J.~Wang, L.~Hui and J.~Khoury, \emph{{No-Go Theorems for Generalized Chameleon
  Field Theories}},
  \href{https://doi.org/10.1103/PhysRevLett.109.241301}{\emph{Phys. Rev. Lett.}
  {\bfseries 109} (2012) 241301}
  [\href{https://arxiv.org/abs/1208.4612}{{\ttfamily 1208.4612}}].

\bibitem{Brax:2012gr}
P.~Brax, A.-C.~Davis, B.~Li and H.A.~Winther, \emph{{A Unified Description of
  Screened Modified Gravity}},
  \href{https://doi.org/10.1103/PhysRevD.86.044015}{\emph{Phys. Rev. D}
  {\bfseries 86} (2012) 044015}
  [\href{https://arxiv.org/abs/1203.4812}{{\ttfamily 1203.4812}}].

\bibitem{Kaneta:2023rby}
K.~Kaneta, K.-y.~Oda and M.~Yoshimura, \emph{{Constraints on extended
  Jordan-Brans-Dicke gravity}},
  \href{https://doi.org/10.1088/1475-7516/2023/10/040}{\emph{JCAP} {\bfseries
  10} (2023) 040} [\href{https://arxiv.org/abs/2304.08656}{{\ttfamily
  2304.08656}}].

\bibitem{Planck:2018vyg}
{\scshape Planck} collaboration, \emph{{Planck 2018 results. VI. Cosmological
  parameters}},
  \href{https://doi.org/10.1051/0004-6361/201833910}{\emph{Astron. Astrophys.}
  {\bfseries 641} (2020) A6}
  [\href{https://arxiv.org/abs/1807.06209}{{\ttfamily 1807.06209}}].

\bibitem{Zlatev:1998tr}
I.~Zlatev, L.-M.~Wang and P.J.~Steinhardt, \emph{{Quintessence, cosmic
  coincidence, and the cosmological constant}},
  \href{https://doi.org/10.1103/PhysRevLett.82.896}{\emph{Phys. Rev. Lett.}
  {\bfseries 82} (1999) 896}
  [\href{https://arxiv.org/abs/astro-ph/9807002}{{\ttfamily
  astro-ph/9807002}}].

\bibitem{Kaneta:2022kjj}
K.~Kaneta, H.-S.~Lee, J.~Lee and J.~Yi, \emph{{Gauged quintessence}},
  \href{https://doi.org/10.1088/1475-7516/2023/02/005}{\emph{JCAP} {\bfseries
  02} (2023) 005} [\href{https://arxiv.org/abs/2208.09229}{{\ttfamily
  2208.09229}}].

\bibitem{Kaneta:2023lki}
K.~Kaneta, H.-S.~Lee, J.~Lee and J.~Yi, \emph{{Misalignment mechanism for a
  mass-varying vector boson}},
  \href{https://doi.org/10.1088/1475-7516/2023/09/017}{\emph{JCAP} {\bfseries
  09} (2023) 017} [\href{https://arxiv.org/abs/2306.01291}{{\ttfamily
  2306.01291}}].

\bibitem{Banerjee:2020xcn}
A.~Banerjee, H.~Cai, L.~Heisenberg, E.O.~Colg\'ain, M.M.~Sheikh-Jabbari and
  T.~Yang, \emph{{Hubble sinks in the low-redshift swampland}},
  \href{https://doi.org/10.1103/PhysRevD.103.L081305}{\emph{Phys. Rev. D}
  {\bfseries 103} (2021) L081305}
  [\href{https://arxiv.org/abs/2006.00244}{{\ttfamily 2006.00244}}].

\bibitem{Lee:2022cyh}
B.-H.~Lee, W.~Lee, E.O.~Colg\'ain, M.M.~Sheikh-Jabbari and S.~Thakur, \emph{{Is
  local H $_{0}$ at odds with dark energy EFT?}},
  \href{https://doi.org/10.1088/1475-7516/2022/04/004}{\emph{JCAP} {\bfseries
  04} (2022) 004} [\href{https://arxiv.org/abs/2202.03906}{{\ttfamily
  2202.03906}}].

\bibitem{Nelson:2011sf}
A.E.~Nelson and J.~Scholtz, \emph{{Dark Light, Dark Matter and the Misalignment
  Mechanism}}, \href{https://doi.org/10.1103/PhysRevD.84.103501}{\emph{Phys.
  Rev. D} {\bfseries 84} (2011) 103501}
  [\href{https://arxiv.org/abs/1105.2812}{{\ttfamily 1105.2812}}].

\bibitem{PhysRevLett.81.3067}
S.M.~Carroll, \emph{Quintessence and the rest of the world: Suppressing
  long-range interactions},
  \href{https://doi.org/10.1103/PhysRevLett.81.3067}{\emph{Phys. Rev. Lett.}
  {\bfseries 81} (1998) 3067}.

\bibitem{Shaw:2005ip}
D.J.~Shaw, \emph{{Charge non-conservation, dequantisation, and induced electric
  dipole moments in varying-alpha theories}},
  \href{https://doi.org/10.1016/j.physletb.2005.10.033}{\emph{Phys. Lett. B}
  {\bfseries 632} (2006) 105}
  [\href{https://arxiv.org/abs/hep-th/0509093}{{\ttfamily hep-th/0509093}}].

\bibitem{Ferlito:2022mok}
F.~Ferlito, S.~Vagnozzi, D.F.~Mota and M.~Baldi, \emph{{Cosmological direct
  detection of dark energy: Non-linear structure formation signatures of dark
  energy scattering with visible matter}},
  \href{https://doi.org/10.1093/mnras/stac649}{\emph{Mon. Not. Roy. Astron.
  Soc.} {\bfseries 512} (2022) 1885}
  [\href{https://arxiv.org/abs/2201.04528}{{\ttfamily 2201.04528}}].

\bibitem{Vagnozzi:2021quy}
S.~Vagnozzi, L.~Visinelli, P.~Brax, A.-C.~Davis and J.~Sakstein, \emph{{Direct
  detection of dark energy: The XENON1T excess and future prospects}},
  \href{https://doi.org/10.1103/PhysRevD.104.063023}{\emph{Phys. Rev. D}
  {\bfseries 104} (2021) 063023}
  [\href{https://arxiv.org/abs/2103.15834}{{\ttfamily 2103.15834}}].

\bibitem{Berghaus:2020ekh}
K.V.~Berghaus, P.W.~Graham, D.E.~Kaplan, G.D.~Moore and S.~Rajendran,
  \emph{{Dark energy radiation}},
  \href{https://doi.org/10.1103/PhysRevD.104.083520}{\emph{Phys. Rev. D}
  {\bfseries 104} (2021) 083520}
  [\href{https://arxiv.org/abs/2012.10549}{{\ttfamily 2012.10549}}].

\bibitem{Berghaus:2023ypi}
K.V.~Berghaus, T.~Karwal, V.~Miranda and T.~Brinckmann, \emph{{The Cosmology of
  Dark Energy Radiation}},  \href{https://arxiv.org/abs/2311.08638}{{\ttfamily
  2311.08638}}.

\bibitem{Brax:2007ak}
P.~Brax, C.~van~de Bruck and A.-C.~Davis, \emph{{Compatibility of the
  chameleon-field model with fifth-force experiments, cosmology, and PVLAS and
  CAST results}},
  \href{https://doi.org/10.1103/PhysRevLett.99.121103}{\emph{Phys. Rev. Lett.}
  {\bfseries 99} (2007) 121103}
  [\href{https://arxiv.org/abs/hep-ph/0703243}{{\ttfamily hep-ph/0703243}}].

\bibitem{Brax:2007hi}
P.~Brax, C.~van~de Bruck, A.-C.~Davis, D.F.~Mota and D.J.~Shaw, \emph{{Testing
  Chameleon Theories with Light Propagating through a Magnetic Field}},
  \href{https://doi.org/10.1103/PhysRevD.76.085010}{\emph{Phys. Rev. D}
  {\bfseries 76} (2007) 085010}
  [\href{https://arxiv.org/abs/0707.2801}{{\ttfamily 0707.2801}}].

\bibitem{Homma:2019rqb}
K.~Homma and Y.~Kirita, \emph{{Stimulated radar collider for probing
  gravitationally weak coupling pseudo Nambu-Goldstone bosons}},
  \href{https://doi.org/10.1007/JHEP09(2020)095}{\emph{JHEP} {\bfseries 09}
  (2020) 095} [\href{https://arxiv.org/abs/1909.00983}{{\ttfamily
  1909.00983}}].

\bibitem{Vagnozzi:2019kvw}
S.~Vagnozzi, L.~Visinelli, O.~Mena and D.F.~Mota, \emph{{Do we have any hope of
  detecting scattering between dark energy and baryons through cosmology?}},
  \href{https://doi.org/10.1093/mnras/staa311}{\emph{Mon. Not. Roy. Astron.
  Soc.} {\bfseries 493} (2020) 1139}
  [\href{https://arxiv.org/abs/1911.12374}{{\ttfamily 1911.12374}}].

\bibitem{Brax:2009aw}
P.~Brax, C.~Burrage, A.-C.~Davis, D.~Seery and A.~Weltman, \emph{{Collider
  constraints on interactions of dark energy with the Standard Model}},
  \href{https://doi.org/10.1088/1126-6708/2009/09/128}{\emph{JHEP} {\bfseries
  09} (2009) 128} [\href{https://arxiv.org/abs/0904.3002}{{\ttfamily
  0904.3002}}].

\bibitem{Brax:2009ey}
P.~Brax, C.~Burrage, A.-C.~Davis, D.~Seery and A.~Weltman, \emph{{Higgs
  production as a probe of Chameleon Dark Energy}},
  \href{https://doi.org/10.1103/PhysRevD.81.103524}{\emph{Phys. Rev. D}
  {\bfseries 81} (2010) 103524}
  [\href{https://arxiv.org/abs/0911.1267}{{\ttfamily 0911.1267}}].

\bibitem{Brax:2015hma}
P.~Brax, C.~Burrage and C.~Englert, \emph{{Disformal dark energy at
  colliders}}, \href{https://doi.org/10.1103/PhysRevD.92.044036}{\emph{Phys.
  Rev. D} {\bfseries 92} (2015) 044036}
  [\href{https://arxiv.org/abs/1506.04057}{{\ttfamily 1506.04057}}].

\bibitem{Brax:2016did}
P.~Brax, C.~Burrage, C.~Englert and M.~Spannowsky, \emph{{LHC Signatures Of
  Scalar Dark Energy}},
  \href{https://doi.org/10.1103/PhysRevD.94.084054}{\emph{Phys. Rev. D}
  {\bfseries 94} (2016) 084054}
  [\href{https://arxiv.org/abs/1604.04299}{{\ttfamily 1604.04299}}].

\bibitem{ATLAS:2019wdu}
{\scshape ATLAS} collaboration, \emph{{Constraints on mediator-based dark
  matter and scalar dark energy models using $\sqrt s = 13$ TeV $pp$ collision
  data collected by the ATLAS detector}},
  \href{https://doi.org/10.1007/JHEP05(2019)142}{\emph{JHEP} {\bfseries 05}
  (2019) 142} [\href{https://arxiv.org/abs/1903.01400}{{\ttfamily
  1903.01400}}].

\bibitem{Casas:1991ky}
J.A.~Casas, J.~Garcia-Bellido and M.~Quiros, \emph{{Scalar - tensor theories of
  gravity with phi dependent masses}},
  \href{https://doi.org/10.1088/0264-9381/9/5/018}{\emph{Class. Quant. Grav.}
  {\bfseries 9} (1992) 1371}
  [\href{https://arxiv.org/abs/hep-ph/9204213}{{\ttfamily hep-ph/9204213}}].

\bibitem{Garcia-Bellido:1992xlz}
J.~Garcia-Bellido, \emph{{Dark matter with variable masses}},
  \href{https://doi.org/10.1142/S0218271893000076}{\emph{Int. J. Mod. Phys. D}
  {\bfseries 2} (1993) 85}
  [\href{https://arxiv.org/abs/hep-ph/9205216}{{\ttfamily hep-ph/9205216}}].

\bibitem{Anderson:1997un}
G.W.~Anderson and S.M.~Carroll, \emph{{Dark matter with time dependent mass}},
  in \emph{{1st International Conference on Particle Physics and the Early
  Universe}}, pp.~227--229, 9, 1997,
  \href{https://doi.org/10.1142/9789814447263_0025}{DOI}
  [\href{https://arxiv.org/abs/astro-ph/9711288}{{\ttfamily
  astro-ph/9711288}}].

\bibitem{Fardon:2003eh}
R.~Fardon, A.E.~Nelson and N.~Weiner, \emph{{Dark energy from mass varying
  neutrinos}}, \href{https://doi.org/10.1088/1475-7516/2004/10/005}{\emph{JCAP}
  {\bfseries 10} (2004) 005}
  [\href{https://arxiv.org/abs/astro-ph/0309800}{{\ttfamily
  astro-ph/0309800}}].

\bibitem{Berlin:2016bdv}
A.~Berlin and D.~Hooper, \emph{{Axion-Assisted Production of Sterile Neutrino
  Dark Matter}}, \href{https://doi.org/10.1103/PhysRevD.95.075017}{\emph{Phys.
  Rev. D} {\bfseries 95} (2017) 075017}
  [\href{https://arxiv.org/abs/1610.03849}{{\ttfamily 1610.03849}}].

\bibitem{Krnjaic:2017zlz}
G.~Krnjaic, P.A.N.~Machado and L.~Necib, \emph{{Distorted neutrino oscillations
  from time varying cosmic fields}},
  \href{https://doi.org/10.1103/PhysRevD.97.075017}{\emph{Phys. Rev. D}
  {\bfseries 97} (2018) 075017}
  [\href{https://arxiv.org/abs/1705.06740}{{\ttfamily 1705.06740}}].

\bibitem{Davoudiasl:2019xeb}
H.~Davoudiasl and G.~Mohlabeng, \emph{{Getting a THUMP from a WIMP}},
  \href{https://doi.org/10.1007/JHEP04(2020)177}{\emph{JHEP} {\bfseries 04}
  (2020) 177} [\href{https://arxiv.org/abs/1912.05572}{{\ttfamily
  1912.05572}}].

\bibitem{Boubekeur:2023fqo}
L.~Boubekeur and S.~Profumo, \emph{{Tremaine-Gunn limit with mass-varying
  particles}}, \href{https://doi.org/10.1103/PhysRevD.107.103535}{\emph{Phys.
  Rev. D} {\bfseries 107} (2023) 103535}
  [\href{https://arxiv.org/abs/2302.10246}{{\ttfamily 2302.10246}}].

\bibitem{ChoeJo:2023ffp}
Y.~ChoeJo, Y.~Kim and H.-S.~Lee, \emph{{Dirac-Majorana neutrino type
  oscillation induced by a wave dark matter}},
  \href{https://doi.org/10.1103/PhysRevD.108.095028}{\emph{Phys. Rev. D}
  {\bfseries 108} (2023) 095028}
  [\href{https://arxiv.org/abs/2305.16900}{{\ttfamily 2305.16900}}].

\bibitem{ChoeJo:2023cnx}
Y.~ChoeJo, K.~Enomoto, Y.~Kim and H.-S.~Lee, \emph{{Second leptogenesis:
  Unraveling the baryon-lepton asymmetry discrepancy}},
  \href{https://arxiv.org/abs/2311.16672}{{\ttfamily 2311.16672}}.

\bibitem{Holdom:1985ag}
B.~Holdom, \emph{{Two U(1)'s and Epsilon Charge Shifts}},
  \href{https://doi.org/10.1016/0370-2693(86)91377-8}{\emph{Phys. Lett. B}
  {\bfseries 166} (1986) 196}.

\bibitem{Altherr:1992mf}
T.~Altherr and U.~Kraemmer, \emph{{Gauge field theory methods for
  ultradegenerate and ultrarelativistic plasmas}},
  \href{https://doi.org/10.1016/0927-6505(92)90014-Q}{\emph{Astropart. Phys.}
  {\bfseries 1} (1992) 133}.

\bibitem{Braaten:1993jw}
E.~Braaten and D.~Segel, \emph{{Neutrino energy loss from the plasma process at
  all temperatures and densities}},
  \href{https://doi.org/10.1103/PhysRevD.48.1478}{\emph{Phys. Rev. D}
  {\bfseries 48} (1993) 1478}
  [\href{https://arxiv.org/abs/hep-ph/9302213}{{\ttfamily hep-ph/9302213}}].

\bibitem{Redondo:2008aa}
J.~Redondo, \emph{{Helioscope Bounds on Hidden Sector Photons}},
  \href{https://doi.org/10.1088/1475-7516/2008/07/008}{\emph{JCAP} {\bfseries
  07} (2008) 008} [\href{https://arxiv.org/abs/0801.1527}{{\ttfamily
  0801.1527}}].

\bibitem{Redondo:2008ec}
J.~Redondo and M.~Postma, \emph{{Massive hidden photons as lukewarm dark
  matter}}, \href{https://doi.org/10.1088/1475-7516/2009/02/005}{\emph{JCAP}
  {\bfseries 02} (2009) 005} [\href{https://arxiv.org/abs/0811.0326}{{\ttfamily
  0811.0326}}].

\bibitem{Steinhardt:1999nw}
P.J.~Steinhardt, L.-M.~Wang and I.~Zlatev, \emph{{Cosmological tracking
  solutions}}, \href{https://doi.org/10.1103/PhysRevD.59.123504}{\emph{Phys.
  Rev. D} {\bfseries 59} (1999) 123504}
  [\href{https://arxiv.org/abs/astro-ph/9812313}{{\ttfamily
  astro-ph/9812313}}].

\bibitem{Linde:1978px}
A.D.~Linde, \emph{{Phase Transitions in Gauge Theories and Cosmology}},
  \href{https://doi.org/10.1088/0034-4885/42/3/001}{\emph{Rept. Prog. Phys.}
  {\bfseries 42} (1979) 389}.

\bibitem{Kaneta:2016wvf}
K.~Kaneta, H.-S.~Lee and S.~Yun, \emph{{Portal Connecting Dark Photons and
  Axions}}, \href{https://doi.org/10.1103/PhysRevLett.118.101802}{\emph{Phys.
  Rev. Lett.} {\bfseries 118} (2017) 101802}
  [\href{https://arxiv.org/abs/1611.01466}{{\ttfamily 1611.01466}}].

\bibitem{Kaneta:2017wfh}
K.~Kaneta, H.-S.~Lee and S.~Yun, \emph{{Dark photon relic dark matter
  production through the dark axion portal}},
  \href{https://doi.org/10.1103/PhysRevD.95.115032}{\emph{Phys. Rev. D}
  {\bfseries 95} (2017) 115032}
  [\href{https://arxiv.org/abs/1704.07542}{{\ttfamily 1704.07542}}].

\bibitem{EscuderoAbenza:2020cmq}
M.~Escudero~Abenza, \emph{{Precision early universe thermodynamics made simple:
  $N_{\rm eff}$ and neutrino decoupling in the Standard Model and beyond}},
  \href{https://doi.org/10.1088/1475-7516/2020/05/048}{\emph{JCAP} {\bfseries
  05} (2020) 048} [\href{https://arxiv.org/abs/2001.04466}{{\ttfamily
  2001.04466}}].

\bibitem{Thomas:2019ran}
L.C.~Thomas, T.~Dezen, E.B.~Grohs and C.T.~Kishimoto, \emph{{Electron-Positron
  Annihilation Freeze-Out in the Early Universe}},
  \href{https://doi.org/10.1103/PhysRevD.101.063507}{\emph{Phys. Rev. D}
  {\bfseries 101} (2020) 063507}
  [\href{https://arxiv.org/abs/1910.14050}{{\ttfamily 1910.14050}}].

\bibitem{Ibe:2019gpv}
M.~Ibe, S.~Kobayashi, Y.~Nakayama and S.~Shirai, \emph{{Cosmological constraint
  on dark photon from N$_{eff}$}},
  \href{https://doi.org/10.1007/JHEP04(2020)009}{\emph{JHEP} {\bfseries 04}
  (2020) 009} [\href{https://arxiv.org/abs/1912.12152}{{\ttfamily
  1912.12152}}].

\bibitem{Pospelov:2008jk}
M.~Pospelov, A.~Ritz and M.B.~Voloshin, \emph{{Bosonic super-WIMPs as keV-scale
  dark matter}}, \href{https://doi.org/10.1103/PhysRevD.78.115012}{\emph{Phys.
  Rev. D} {\bfseries 78} (2008) 115012}
  [\href{https://arxiv.org/abs/0807.3279}{{\ttfamily 0807.3279}}].

\bibitem{McDermott:2017qcg}
S.D.~McDermott, H.H.~Patel and H.~Ramani, \emph{{Dark Photon Decay Beyond The
  Euler-Heisenberg Limit}},
  \href{https://doi.org/10.1103/PhysRevD.97.073005}{\emph{Phys. Rev. D}
  {\bfseries 97} (2018) 073005}
  [\href{https://arxiv.org/abs/1705.00619}{{\ttfamily 1705.00619}}].

\bibitem{Jaeckel:2008fi}
J.~Jaeckel, J.~Redondo and A.~Ringwald, \emph{{Signatures of a hidden cosmic
  microwave background}},
  \href{https://doi.org/10.1103/PhysRevLett.101.131801}{\emph{Phys. Rev. Lett.}
  {\bfseries 101} (2008) 131801}
  [\href{https://arxiv.org/abs/0804.4157}{{\ttfamily 0804.4157}}].

\bibitem{Bjorken:2009mm}
J.D.~Bjorken, R.~Essig, P.~Schuster and N.~Toro, \emph{{New Fixed-Target
  Experiments to Search for Dark Gauge Forces}},
  \href{https://doi.org/10.1103/PhysRevD.80.075018}{\emph{Phys. Rev. D}
  {\bfseries 80} (2009) 075018}
  [\href{https://arxiv.org/abs/0906.0580}{{\ttfamily 0906.0580}}].

\bibitem{Chang:2016ntp}
J.H.~Chang, R.~Essig and S.D.~McDermott, \emph{{Revisiting Supernova 1987A
  Constraints on Dark Photons}},
  \href{https://doi.org/10.1007/JHEP01(2017)107}{\emph{JHEP} {\bfseries 01}
  (2017) 107} [\href{https://arxiv.org/abs/1611.03864}{{\ttfamily
  1611.03864}}].

\bibitem{DeRocco:2019njg}
W.~DeRocco, P.W.~Graham, D.~Kasen, G.~Marques-Tavares and S.~Rajendran,
  \emph{{Observable signatures of dark photons from supernovae}},
  \href{https://doi.org/10.1007/JHEP02(2019)171}{\emph{JHEP} {\bfseries 02}
  (2019) 171} [\href{https://arxiv.org/abs/1901.08596}{{\ttfamily
  1901.08596}}].

\bibitem{Arkani-Hamed:2006emk}
N.~Arkani-Hamed, L.~Motl, A.~Nicolis and C.~Vafa, \emph{{The String landscape,
  black holes and gravity as the weakest force}},
  \href{https://doi.org/10.1088/1126-6708/2007/06/060}{\emph{JHEP} {\bfseries
  06} (2007) 060} [\href{https://arxiv.org/abs/hep-th/0601001}{{\ttfamily
  hep-th/0601001}}].

\bibitem{Heidenreich:2016aqi}
B.~Heidenreich, M.~Reece and T.~Rudelius, \emph{{Evidence for a sublattice weak
  gravity conjecture}},
  \href{https://doi.org/10.1007/JHEP08(2017)025}{\emph{JHEP} {\bfseries 08}
  (2017) 025} [\href{https://arxiv.org/abs/1606.08437}{{\ttfamily
  1606.08437}}].

\bibitem{Heidenreich:2017sim}
B.~Heidenreich, M.~Reece and T.~Rudelius, \emph{{The Weak Gravity Conjecture
  and Emergence from an Ultraviolet Cutoff}},
  \href{https://doi.org/10.1140/epjc/s10052-018-5811-3}{\emph{Eur. Phys. J. C}
  {\bfseries 78} (2018) 337}
  [\href{https://arxiv.org/abs/1712.01868}{{\ttfamily 1712.01868}}].

\bibitem{Gherghetta:2019coi}
T.~Gherghetta, J.~Kersten, K.~Olive and M.~Pospelov, \emph{{Evaluating the
  price of tiny kinetic mixing}},
  \href{https://doi.org/10.1103/PhysRevD.100.095001}{\emph{Phys. Rev. D}
  {\bfseries 100} (2019) 095001}
  [\href{https://arxiv.org/abs/1909.00696}{{\ttfamily 1909.00696}}].

\bibitem{Hu:1992dc}
W.~Hu and J.~Silk, \emph{{Thermalization and spectral distortions of the cosmic
  background radiation}},
  \href{https://doi.org/10.1103/PhysRevD.48.485}{\emph{Phys. Rev. D} {\bfseries
  48} (1993) 485}.

\bibitem{Fixsen:1996nj}
D.J.~Fixsen, E.S.~Cheng, J.M.~Gales, J.C.~Mather, R.A.~Shafer and E.L.~Wright,
  \emph{{The Cosmic Microwave Background spectrum from the full COBE FIRAS data
  set}}, \href{https://doi.org/10.1086/178173}{\emph{Astrophys. J.} {\bfseries
  473} (1996) 576} [\href{https://arxiv.org/abs/astro-ph/9605054}{{\ttfamily
  astro-ph/9605054}}].

\bibitem{Bianchini:2022dqh}
F.~Bianchini and G.~Fabbian, \emph{{CMB spectral distortions revisited: A new
  take on \ensuremath{\mu} distortions and primordial non-Gaussianities from
  FIRAS data}}, \href{https://doi.org/10.1103/PhysRevD.106.063527}{\emph{Phys.
  Rev. D} {\bfseries 106} (2022) 063527}
  [\href{https://arxiv.org/abs/2206.02762}{{\ttfamily 2206.02762}}].

\bibitem{Chluba:2013vsa}
J.~Chluba, \emph{{Green's function of the cosmological thermalization
  problem}}, \href{https://doi.org/10.1093/mnras/stt1025}{\emph{Mon. Not. Roy.
  Astron. Soc.} {\bfseries 434} (2013) 352}
  [\href{https://arxiv.org/abs/1304.6120}{{\ttfamily 1304.6120}}].

\bibitem{Chluba:2013pya}
J.~Chluba and D.~Jeong, \emph{{Teasing bits of information out of the CMB
  energy spectrum}}, \href{https://doi.org/10.1093/mnras/stt2327}{\emph{Mon.
  Not. Roy. Astron. Soc.} {\bfseries 438} (2014) 2065}
  [\href{https://arxiv.org/abs/1306.5751}{{\ttfamily 1306.5751}}].

\bibitem{Chluba:2015hma}
J.~Chluba, \emph{{Green's function of the cosmological thermalization problem
  \textendash{} II. Effect of photon injection and constraints}},
  \href{https://doi.org/10.1093/mnras/stv2243}{\emph{Mon. Not. Roy. Astron.
  Soc.} {\bfseries 454} (2015) 4182}
  [\href{https://arxiv.org/abs/1506.06582}{{\ttfamily 1506.06582}}].

\bibitem{Chluba:2016bvg}
J.~Chluba, \emph{{Which spectral distortions does $\Lambda$CDM actually
  predict?}}, \href{https://doi.org/10.1093/mnras/stw945}{\emph{Mon. Not. Roy.
  Astron. Soc.} {\bfseries 460} (2016) 227}
  [\href{https://arxiv.org/abs/1603.02496}{{\ttfamily 1603.02496}}].

\bibitem{gendreau1995asca}
K.C.~Gendreau, R.~Mushotzky, A.C.~Fabian, S.S.~Holt, T.~Kii, P.J.~Serlemitsos
  et~al., \emph{Asca observations of the spectrum of the x-ray background},
  {\emph{PASJ} {\bfseries 47} (1995) L5}.

\bibitem{kappadath1998measurement}
S.C.~Kappadath, \emph{Measurement of the cosmic diffuse gamma-ray spectrum from
  800 kev to 30 mev}, {\emph{Ph. D. Thesis} (1998) 4873}.

\bibitem{Gruber:1999yr}
D.E.~Gruber, J.L.~Matteson, L.E.~Peterson and G.V.~Jung, \emph{{The spectrum of
  diffuse cosmic hard x-rays measured with heao-1}},
  \href{https://doi.org/10.1086/307450}{\emph{Astrophys. J.} {\bfseries 520}
  (1999) 124} [\href{https://arxiv.org/abs/astro-ph/9903492}{{\ttfamily
  astro-ph/9903492}}].

\bibitem{Strong:2004de}
A.W.~Strong, I.V.~Moskalenko and O.~Reimer, \emph{{Diffuse galactic continuum
  gamma rays. A Model compatible with EGRET data and cosmic-ray measurements}},
  \href{https://doi.org/10.1086/423193}{\emph{Astrophys. J.} {\bfseries 613}
  (2004) 962} [\href{https://arxiv.org/abs/astro-ph/0406254}{{\ttfamily
  astro-ph/0406254}}].

\bibitem{Bouchet:2008rp}
L.~Bouchet, E.~Jourdain, J.P.~Roques, A.~Strong, R.~Diehl, F.~Lebrun et~al.,
  \emph{{INTEGRAL SPI All-Sky View in Soft Gamma Rays: Study of Point Source
  and Galactic Diffuse Emissions}},
  \href{https://doi.org/10.1086/529489}{\emph{Astrophys. J.} {\bfseries 679}
  (2008) 1315} [\href{https://arxiv.org/abs/0801.2086}{{\ttfamily 0801.2086}}].

\bibitem{Ajello:2008xb}
M.~Ajello et~al., \emph{{Cosmic X-ray background and Earth albedo Spectra with
  Swift/BAT}}, \href{https://doi.org/10.1086/592595}{\emph{Astrophys. J.}
  {\bfseries 689} (2008) 666}
  [\href{https://arxiv.org/abs/0808.3377}{{\ttfamily 0808.3377}}].

\bibitem{Ackermann2012}
M.~Ackermann, M.~Ajello, W.B.~Atwood, L.~Baldini, J.~Ballet, G.~Barbiellini
  et~al., \emph{Fermi-lat observations of the diffuse $\gamma$-ray emission:
  Implications for cosmic rays and the interstellar medium},
  \href{https://doi.org/10.1088/0004-637x/750/1/3}{\emph{Astrophys. J.}
  {\bfseries 750} (2012) 3}.

\bibitem{Masso:1999wj}
E.~Masso and R.~Toldra, \emph{{Photon spectrum produced by the late decay of a
  cosmic neutrino background}},
  \href{https://doi.org/10.1103/PhysRevD.60.083503}{\emph{Phys. Rev. D}
  {\bfseries 60} (1999) 083503}
  [\href{https://arxiv.org/abs/astro-ph/9903397}{{\ttfamily
  astro-ph/9903397}}].

\bibitem{Abazajian:2001vt}
K.~Abazajian, G.M.~Fuller and W.H.~Tucker, \emph{{Direct detection of warm dark
  matter in the X-ray}}, \href{https://doi.org/10.1086/323867}{\emph{Astrophys.
  J.} {\bfseries 562} (2001) 593}
  [\href{https://arxiv.org/abs/astro-ph/0106002}{{\ttfamily
  astro-ph/0106002}}].

\bibitem{Boyarsky:2005us}
A.~Boyarsky, A.~Neronov, O.~Ruchayskiy and M.~Shaposhnikov, \emph{{Constraints
  on sterile neutrino as a dark matter candidate from the diffuse x-ray
  background}},
  \href{https://doi.org/10.1111/j.1365-2966.2006.10458.x}{\emph{Mon. Not. Roy.
  Astron. Soc.} {\bfseries 370} (2006) 213}
  [\href{https://arxiv.org/abs/astro-ph/0512509}{{\ttfamily
  astro-ph/0512509}}].

\bibitem{Bertone:2007aw}
G.~Bertone, W.~Buchmuller, L.~Covi and A.~Ibarra, \emph{{Gamma-Rays from
  Decaying Dark Matter}},
  \href{https://doi.org/10.1088/1475-7516/2007/11/003}{\emph{JCAP} {\bfseries
  11} (2007) 003} [\href{https://arxiv.org/abs/0709.2299}{{\ttfamily
  0709.2299}}].

\bibitem{Essig:2013goa}
R.~Essig, E.~Kuflik, S.D.~McDermott, T.~Volansky and K.M.~Zurek,
  \emph{{Constraining Light Dark Matter with Diffuse X-Ray and Gamma-Ray
  Observations}}, \href{https://doi.org/10.1007/JHEP11(2013)193}{\emph{JHEP}
  {\bfseries 11} (2013) 193} [\href{https://arxiv.org/abs/1309.4091}{{\ttfamily
  1309.4091}}].

\bibitem{An:2014twa}
H.~An, M.~Pospelov, J.~Pradler and A.~Ritz, \emph{{Direct Detection Constraints
  on Dark Photon Dark Matter}},
  \href{https://doi.org/10.1016/j.physletb.2015.06.018}{\emph{Phys. Lett. B}
  {\bfseries 747} (2015) 331}
  [\href{https://arxiv.org/abs/1412.8378}{{\ttfamily 1412.8378}}].

\bibitem{Yuksel:2007dr}
H.~Yuksel and M.D.~Kistler, \emph{{Circumscribing late dark matter decays model
  independently}},
  \href{https://doi.org/10.1103/PhysRevD.78.023502}{\emph{Phys. Rev. D}
  {\bfseries 78} (2008) 023502}
  [\href{https://arxiv.org/abs/0711.2906}{{\ttfamily 0711.2906}}].

\bibitem{PAMELA:2008gwm}
{\scshape PAMELA} collaboration, \emph{{An anomalous positron abundance in
  cosmic rays with energies 1.5-100 GeV}},
  \href{https://doi.org/10.1038/nature07942}{\emph{Nature} {\bfseries 458}
  (2009) 607} [\href{https://arxiv.org/abs/0810.4995}{{\ttfamily 0810.4995}}].

\bibitem{Fermi-LAT:2011baq}
{\scshape Fermi-LAT} collaboration, \emph{{Measurement of separate cosmic-ray
  electron and positron spectra with the Fermi Large Area Telescope}},
  \href{https://doi.org/10.1103/PhysRevLett.108.011103}{\emph{Phys. Rev. Lett.}
  {\bfseries 108} (2012) 011103}
  [\href{https://arxiv.org/abs/1109.0521}{{\ttfamily 1109.0521}}].

\bibitem{AMS:2013fma}
{\scshape AMS} collaboration, \emph{{First Result from the Alpha Magnetic
  Spectrometer on the International Space Station: Precision Measurement of the
  Positron Fraction in Primary Cosmic Rays of 0.5\textendash{}350 GeV}},
  \href{https://doi.org/10.1103/PhysRevLett.110.141102}{\emph{Phys. Rev. Lett.}
  {\bfseries 110} (2013) 141102}.

\bibitem{HESS:2017clw}
{\scshape H.E.S.S.} collaboration, \emph{{Contributions of the High Energy
  Stereoscopic System (H.E.S.S.) to the 35th International Cosmic Ray
  Conference (ICRC), Busan, Korea}},  9, 2017
  [\href{https://arxiv.org/abs/1709.06442}{{\ttfamily 1709.06442}}].

\bibitem{DAMPE:2017fbg}
{\scshape DAMPE} collaboration, \emph{{Direct detection of a break in the
  teraelectronvolt cosmic-ray spectrum of electrons and positrons}},
  \href{https://doi.org/10.1038/nature24475}{\emph{Nature} {\bfseries 552}
  (2017) 63} [\href{https://arxiv.org/abs/1711.10981}{{\ttfamily 1711.10981}}].

\bibitem{CALET:2017uxd}
{\scshape CALET} collaboration, \emph{{Energy Spectrum of Cosmic-Ray Electron
  and Positron from 10 GeV to 3 TeV Observed with the Calorimetric Electron
  Telescope on the International Space Station}},
  \href{https://doi.org/10.1103/PhysRevLett.119.181101}{\emph{Phys. Rev. Lett.}
  {\bfseries 119} (2017) 181101}
  [\href{https://arxiv.org/abs/1712.01711}{{\ttfamily 1712.01711}}].

\bibitem{Hillas1984}
A.~Hillas, \emph{The origin of ultra-high-energy cosmic rays},
  \href{https://doi.org/10.1146/annurev.aa.22.090184.002233}{\emph{Anun. Rev.
  Astron.} {\bfseries 22} (1984) 425}
  [\href{https://arxiv.org/abs/https://doi.org/10.1146/annurev.aa.22.090184.002233}{{\ttfamily
  https://doi.org/10.1146/annurev.aa.22.090184.002233}}].

\bibitem{Davoudiasl:2012ag}
H.~Davoudiasl, H.-S.~Lee and W.J.~Marciano, \emph{{'Dark' Z implications for
  Parity Violation, Rare Meson Decays, and Higgs Physics}},
  \href{https://doi.org/10.1103/PhysRevD.85.115019}{\emph{Phys. Rev. D}
  {\bfseries 85} (2012) 115019}
  [\href{https://arxiv.org/abs/1203.2947}{{\ttfamily 1203.2947}}].

\bibitem{Davoudiasl:2014kua}
H.~Davoudiasl, H.-S.~Lee and W.J.~Marciano, \emph{{Muon $g - 2$, rare kaon
  decays, and parity violation from dark bosons}},
  \href{https://doi.org/10.1103/PhysRevD.89.095006}{\emph{Phys. Rev. D}
  {\bfseries 89} (2014) 095006}
  [\href{https://arxiv.org/abs/1402.3620}{{\ttfamily 1402.3620}}].

\bibitem{Davoudiasl:2023cnc}
H.~Davoudiasl, K.~Enomoto, H.-S.~Lee, J.~Lee and W.J.~Marciano,
  \emph{{Searching for new physics effects in future W mass and
  $sin^2\ensuremath{\theta_W}(Q^2)$ determinations}},
  \href{https://doi.org/10.1103/PhysRevD.108.115018}{\emph{Phys. Rev. D}
  {\bfseries 108} (2023) 115018}
  [\href{https://arxiv.org/abs/2309.04060}{{\ttfamily 2309.04060}}].

\end{thebibliography}\endgroup

\end{document}